\documentclass[12pt]{article}
\usepackage{latexsym}  
\def\Journal#1#2#3#4{{#1} {\bf #2} (#4) #3}

\def\NPB{{\em Nucl. Phys.} B}
\def\PLB{{\em Phys. Lett.}  B}

\def\IJMPA{{\em Int. J. Mod. Phys.}A}
\def\CMP{\em Comm. Math. Phys.}
\def\CQG{\em Class. Quantum Grav.}
\newenvironment{proof}{\par\noindent\mbox{\bf Proof.}\\}
                      {\QED}
\newtheorem{property}{Property}[section]
\newtheorem{theorem}{Theorem}[section]

\newtheorem{lemma}[theorem]{Lemma}

\newtheorem{gevolg}[theorem]{Corollary}

\newcommand{\QED}{{\hspace*{\fill}\rule{2mm}{2mm}}}
\newcommand{\ft}[2]{{\textstyle\frac{#1}{#2}}}
\newsavebox{\uuunit}
\sbox{\uuunit}
    {\setlength{\unitlength}{0.825em}
     \begin{picture}(0.6,0.7)
        \thinlines
        \put(0,0){\line(1,0){0.5}}
        \put(0.15,0){\line(0,1){0.7}}
        \put(0.35,0){\line(0,1){0.8}}
       \multiput(0.3,0.8)(-0.04,-0.02){12}{\rule{0.5pt}{0.5pt}}
     \end {picture}}
\newcommand {\unity}{\mathord{\!\usebox{\uuunit}}}
\newcommand{\Ka}{K\"ahler}

\newcommand  {\Rbar} {{\mbox{\rm$\mbox{I}\!\mbox{R}$}}}

\newcommand {\Cbar}
    {\mathord{\setlength{\unitlength}{1em}
     \begin{picture}(0.6,0.7)(-0.1,0)
        \put(-0.1,0){\rm C}
        \thicklines
        \put(0.2,0.05){\line(0,1){0.55}}
     \end {picture}}}
\newsavebox{\zzzbar}
\sbox{\zzzbar}
  {\setlength{\unitlength}{0.9em}
  \begin{picture}(0.6,0.7)
  \thinlines
  \put(0,0){\line(1,0){0.6}}
  \put(0,0.75){\line(1,0){0.575}}
  \multiput(0,0)(0.0125,0.025){30}{\rule{0.3pt}{0.3pt}}
  \multiput(0.2,0)(0.0125,0.025){30}{\rule{0.3pt}{0.3pt}}
  \put(0,0.75){\line(0,-1){0.15}}
  \put(0.015,0.75){\line(0,-1){0.1}}
  \put(0.03,0.75){\line(0,-1){0.075}}
  \put(0.045,0.75){\line(0,-1){0.05}}
  \put(0.05,0.75){\line(0,-1){0.025}}
  \put(0.6,0){\line(0,1){0.15}}
  \put(0.585,0){\line(0,1){0.1}}
  \put(0.57,0){\line(0,1){0.075}}
  \put(0.555,0){\line(0,1){0.05}}
  \put(0.55,0){\line(0,1){0.025}}
  \end{picture}}
\newcommand{\Zbar}{\mathord{\!{\usebox{\zzzbar}}}}
\def\IP{\relax{\rm I\kern-.18em P}}
\def\Im{{\rm Im ~}}
\def\Re{{\rm Re ~}}
\newcommand{\beq}{\begin{equation}}
\newcommand{\eeq}{\end{equation}}
\newcommand{\IM}{{\rm Im~}}
\newcommand{\RE}{{\rm Re~}}
\newcommand{\N}{\mbox{$\cal N$}}
\newcommand{\dee}[1]{\mbox{$\partial_{#1}$}}
\newcommand{\Sp}[1]{\mbox{$Sp\left( #1,\Rbar \right) $}}
\newcommand{\symp}[1]{\mbox{$Sp\left( #1,\Rbar \right) $}}
\newcommand{\sinprod}[2]{\mbox{$\langle #1 , #2 \rangle$}}
\newcommand{\eqn}[1]{(\ref{#1})}
\newcommand{\M}{\mbox{${\cal M}$}}
\newcommand{\matrx}[4]{\left(
  \begin{array}{cc}
    {#1} & {#2} \\
    {#3} & {#4}
  \end{array}\right)}
\newcommand{\rvec}[3]{\mbox{$ #1^{#2},\ldots,#1^{#3} $}}
\newcommand{\Dee}[1]{\mbox{${\cal{D}}_{#1}$}}
\newcommand{\rng}[2]{\mbox{${#1}\ldots{#2}$}}
\newcommand{\XJ}{\mbox{$X^{J}$}}
\newcommand{\za}{\mbox{$z^{\alpha}$}}

\newcommand{\ZK}{\mbox{$Z^{K}$}}
\newcommand{\ZL}{\mbox{$Z^{L}$}}

\newcommand{\mtrx}[4]{\left(
  \begin{array}{cc}
    {#1} & {#2} \\
    {#3} & {#4}
  \end{array}\right)}
\newcommand{\dsl}{\not\!\partial}

\csname @addtoreset\endcsname{equation}{section}

\begin{document}
\begin{titlepage}
\begin{flushright} KUL-TF-97/10 \\ hep-th/9703082
\end{flushright}
\vfill
\begin{center}
{\LARGE\bf What is Special K\"ahler Geometry ?}    \\
\vskip 27.mm  \large
{\bf   Ben Craps $^1$, Frederik Roose, \\[2mm] Walter Troost $^2$
and Antoine Van Proeyen $^3$} \\
\vskip 1cm
{\em Instituut voor theoretische fysica}\\
{\em Katholieke Universiteit Leuven, B-3001 Leuven, Belgium}
\end{center}
\vfill

\begin{center}
{\bf ABSTRACT}
\end{center}
\begin{quote}
  The scalars in vector multiplets of N=2 supersymmetric theories in
  4 dimensions exhibit `special  K\"ahler geometry', related
  to  duality symmetries, due to their coupling to the vectors.
  In the literature there is some confusion on the definition of special
  geometry. We show equivalences of some definitions and give
  examples which show that earlier definitions are not equivalent,
  and are not sufficient to restrict the \Ka\ metric to one that
  occurs in $N=2$ supersymmetry. We treat  the rigid as well as the
  local supersymmetry case. The connection is made to moduli spaces
  of Riemann surfaces and Calabi-Yau 3-folds. The conditions for the
  existence of a prepotential translate to a condition on the choice
  of canonical basis of cycles.
\vfill      \hrule width 5.cm
\vskip 2.mm
{\small
\noindent $^1$ Aspirant FWO, Belgium \\
\noindent $^2$ Onderzoeksleider FWO, Belgium \\
\noindent $^3$ Onderzoeksdirecteur FWO, Belgium }
\end{quote}
\begin{flushleft}
 March 1997
\end{flushleft}
\end{titlepage}

\section{Introduction}
The geometry defined by  the complex scalars in the
coupling of $N=2$ vector multiplets to supergravity~\cite{DWVP}
has been given the name \cite{special} `special \Ka\ geometry'. A
similar geometry in $N=2$ {\em rigid} supersymmetry \cite{PKTN2} has been called
`rigid special \Ka'.

Whereas their origin and use in supersymmetric theories
determines the concept of special \Ka\ manifolds, it is also of interest
to have a description that does
not need the full construction of a supergravity action to describe
this class of manifolds. Among other merits,
through capturing the essence, this description may enable one
to find realisations of special geometry in a setting different from supergravity.
Various authors
\cite{special,CdAF,CDFLL,f0art,prtrquat,fresoriabook,ItalianN2}
have proposed such a description, leading to different proposals for a
general definition of the special \Ka\ geometries.
Some of these were inspired by the specific setting of
subvarieties of the moduli spaces of Riemann surfaces or Calabi-Yau
manifolds, where these special \Ka\ geometries appear.
These definitions are not completely
equivalent. We will propose new definitions,
and prove their equivalence. A short r\'esum\'e of the results was
given in \cite{sc2defsg}, and a long version in \cite{benfredlic}.

The recent upsurge of interest in this question is due to duality.
In four dimensions, duality transformations are transformations
between the field strengths of spin-1 fields. In general,
the kinetic terms of
the vectors may depend on scalar fields. As we will be
concerned finally mainly with the kinetic terms of the scalars, it is
sufficient to consider abelian gauge groups. The generic form of the
kinetic terms of the vector fields is then\footnote{We take metric signature
$(-+++)$ and $\epsilon^{0123}=i$.}
\begin{equation}
{\cal L}_1=
\ft14 (\Im {\cal N}_{IJ}){\cal F}_{\mu\nu}^I
{\cal F}^{\mu\nu J}
-\ft i8 (\Re {\cal N}_{IJ})
\epsilon^{\mu\nu\rho\sigma}{\cal F}_{\mu\nu}^I
{\cal F}_{\rho\sigma}^J\ ,   \label{genL01}
\end{equation}
where $I,J=1, ..., m$, the symmetric matrix ${\cal N}_{IJ}$
 depends on the scalars,
and ${\cal F}_{\mu\nu}^I$ are the field strengths of the vector fields.
The imaginary part of the matrix ${\cal N}$
should be negative definite in order to guarantee the positivity of the
kinetic terms of the vectors. This matrix provides the bridge between
the vector and the scalar sectors of the theory. The duality
transformations of the vectors then impose on the scalar sector a
structure which is central to special \Ka\ geometries.
The most useful definitions of special geometry will use a formalism
in which these duality transformations, which take the form of
symplectic transformations \cite{dual,DWVP,CecFerGir,UCCVarDim},
are manifest.
Such a relation between the scalars and the vectors arises naturally in $N=2$
supersymmetry vector multiplets (or larger $N$ extensions).

In section~\ref{ss:sympl} we recall the general
formalism of symplectic transformations for the vectors in 4
dimensions, clarifying in particular the role of the matrix ${\cal N}$,
and recalling some of its properties.
In section~\ref{ss:rigid} the basic concepts of
couplings of vector multiplets in rigid $N=2$ supersymmetry
are explained.
This leads to a first definition of rigid special geometry,
using straightforwardly
the construction of the supersymmetric action, based on a prepotential%
\footnote{We use the terminology `prepotential' for the function
which leads to a \Ka\ potential. The word `prepotential'  originally
denoted an unconstrained superfield in terms of which $N=2$ vector
superfields can be constructed.
In $N=2$ supersymmetry these were found in \cite{N2prepotentials}.}.
Our treatment of the symplectic structure
in a superspace approach reveals that a compatibility condition between
the symplectic and supersymmetric structure largely short-circuits
the use of an action to obtain the field equations of the theory.
A second definition takes the
symplectic structure as central, and does not make use
of a prepotential. We
will show that the two definitions are equivalent. A third definition,
also starting  with the symplectic vector structure, prepares the
stage for showing how certain subspaces of the moduli spaces
of Riemann surfaces provide examples of rigid special geometries:
we will analyze the requirements on these subspaces
and show how various examples in the literature fit in.

The same structure is repeated in section~\ref{ss:local} for
the local (supergravity) case. The first
definition is close to the original construction of actions for
vector multiplets coupled to $N=2$ supergravity using the tensor
calculus \cite{DWVP}. We will first repeat the essential steps of
this construction of the supergravity action. For a review on another
construction, using rheonomic methods \cite{CdAF}, we refer to
\cite{fresoriabook}.
The second definition gives the symplectic structure the central role,
and is based on the one by Strominger \cite{special}.
We point out that, whereas it is adequate within the setting of Calabi-Yau
moduli spaces, in a more general setting it is incomplete.
To show the equivalence between the
two definitions we pay special attention to the formulations of these
models that do not allow a prepotential \cite{f0art}. These formulations
were all obtained by starting from a formulation {\em with} a prepotential
and then performing a symplectic transformation. A main
result of this paper is in this section: the proof that all models
without a prepotential can be obtained from a symplectically rotated
formulation with a prepotential. Since the manifold defined by the scalars
is invariant under such a symplectic transformation, the two
definitions of special manifolds are then equivalent. Again we
give a third definition, in terms of objects which have a geometric
significance in Calabi-Yau moduli spaces (periods of 3-forms), and
of course we then exhibit how this is realised in the Calabi-Yau
context.

In section~\ref{ss:concl} we repeat and clarify the main
conclusions, and give some final remarks. Many detailed proofs are
given in appendices.

 \section{Symplectic transformations}  \label{ss:sympl}
In this section, we remind the reader of the connection
between symplectic
transformations and duality transformations \cite{dual}. Then we
give some general properties of these transformations. It was
realised recently that many aspects of this formalism can be written
quite generally for any dimension and supersymmetry extension
\cite{UCCVarDim}.

Consider a general action of the form ${\cal L}_1$ in \eqn{genL01}
for abelian spin-1 fields. The field equations for the vectors are
\begin{equation}
0=\frac{\partial{\cal L}}{\partial W_\mu^I}=2\partial_\nu
\frac{\partial{\cal L}}{\partial {\cal F}^I_{\mu\nu}}= 2\partial_\nu
\left( \frac{\partial{\cal L}}{\partial {\cal F}^{+I}_{\mu\nu}}
+ \frac{\partial{\cal L}}{\partial {\cal F}^{-I}_{\mu\nu}}\right)\ ,
\end{equation}
where
${\cal F}^\pm _{\mu\nu}= \ft12 \left( {\cal F}_{\mu\nu}  \pm \ft12
\epsilon_{\mu\nu\rho\sigma}{\cal F}^{\rho\sigma}\right) $. Defining
\begin{equation}
  G_{+I }^{\mu\nu}\equiv 2i\frac{\partial{\cal L}}
  {\partial {\cal F}^{+I }_{\mu\nu}}
\ ;\qquad
  G_{-I }^{\mu\nu}\equiv -2i\frac{\partial{\cal L}}
  {\partial {\cal F}^{-I }_{\mu\nu}}
\ ,\label{defG}
\end{equation}
one finds, for ${\cal L}_1$,
\begin{equation}
  G_{+I }^{\mu\nu} =
{\cal N}_{I J }{\cal F}^{+J \,\mu\nu}
\ ;\qquad
  G_{-I }^{\mu\nu} =
\bar{\cal N}_{I J }{\cal F}^{-J \,\mu\nu}
\ .\label{GL1}
\end{equation}
Observe that the symmetry of ${\cal N}$ has been used in these relations.
In the considerations of duality transformations, the spin-1 fields
are represented by field strengths rather than potentials.
This implies that the field
equations are not the only conditions on the field strengths,
but are to be supplemented with the
equations which express that these field strengths are in fact locally
derivable from a vector, viz. the Bianchi identities.
The complete set of equations for the field strengths can then be written as
\begin{eqnarray}
\partial^\mu \Im {\cal F}^{+I }_{\mu\nu} &=&0\ \ \ \ \ {\rm Bianchi\
identities}\nonumber\\
\partial_\mu \Im G_{+I }^{\mu\nu} &=&0\ \ \ \ \  {\rm Field\ equations}\ .
\label{alleqs}
\end{eqnarray}
This set of equations is invariant under  $GL(2m,\Rbar)$
transformations:
\begin{equation}
\pmatrix{\widetilde{\cal F}^+\cr \widetilde G_+\cr}={\cal S}
\pmatrix{{\cal F}^+\cr G_+\cr} =
\pmatrix{A&B\cr C&D\cr}   \pmatrix{{\cal F}^+\cr G_+\cr}\ . \label{FGsympl}
\end{equation}
The $G_{\mu\nu}$ are related to the ${\cal F}_{\mu\nu}$
as in \eqn{defG}. Now we limit the transformations to those that preserve
such a relation.
The relation between $\tilde
G_{\mu\nu}$ and $\tilde {\cal F}_{\mu\nu}$  is given  by the matrix
\begin{equation}
\tilde{\cal N} = (C + D{\cal N})(A+B{\cal N})^{-1}\ .
\label{tilNN}
\end{equation}
As $\tilde G_{+\mu\nu}$ should be the derivative of a transformed
action to describe the field equations,
this requirement imposes that the matrix $\tilde{\cal N}$ should
be symmetric. For a general ${\cal N}$ this implies (using rescalings of
the field strengths)
\begin{equation}
A^T C-C^T A=0\ \ ,\ \ B^T D- D^T B=0\ \ , \ \  A^T D-C^T B=\unity\ .
\label{symplABCD}\end{equation}
These equations express that
\begin{equation}
{\cal S}\equiv\pmatrix{A&B\cr C&D\cr}
\label{defS}
\end{equation}
is a symplectic matrix:
\begin{equation}
{\cal S}\in Sp(2m,\Rbar)\qquad : \qquad
 {\cal S}^T  \Omega   {\cal S}   =  \Omega  \qquad\mbox{with}\qquad
\Omega=\pmatrix{0&\unity  \cr -\unity  &0\cr}\ . \label{cSinSP2m}
\end{equation}

A $2m$ component column $V$ which transforms under the symplectic
transformations as $\widetilde V = {\cal S}V$ is called a symplectic
vector. The invariant inner product of two
symplectic vectors $V$ and $W$ is
\begin{equation}
 \langle V , W \rangle \equiv V^T \Omega W \ .
\end{equation}

The symplectic transformations thus transform solutions of
\eqn{alleqs} into solutions.
However, they are not invariances of the action.
Indeed, writing
\begin{equation}
{\cal L}_1=
\ft 12 \Im \left( {\cal N}_{IJ}\,{\cal F}_{\mu\nu}^{+I}
 {\cal F}^{+\mu\nu\,J }\right) = \ft 12 \Im \left(
 {\cal F}^{+I}   G_{+I} \right) \ ,
\end{equation}
we obtain
\begin{equation}                 \label{action}
\Im \widetilde{\cal F}^{+I}  \widetilde G_{+I}   =
  \Im \left({\cal F}^{+}  G_{+ }\right)
+\Im \left(2 {\cal F}^{+} (C^T B) G_+
 + {\cal F}^{+ }(C^T A) {\cal F}^{+}
 +G_{+ } (D^T B) G_{+ } \right)\ .
\end{equation}
If $C\neq 0, B=0$ the Lagrangian is invariant up to a four--divergence,
as $\Im {\cal F}^+{\cal F}^+=-\ft i4 \epsilon^{\mu\nu\rho\sigma}
{\cal F}_{\mu\nu}{\cal F}_{\rho\sigma}$ and the matrices $A$ and $C$ are real.
For $B\neq 0$ the Lagrangian is {\em not} invariant.

If sources are added to the equations \eqn{alleqs},
the transformations  can be extended to the dyonic solutions of the field
equations by letting the magnetic and electric charges
$\pmatrix{q^I_m\cr q_{e\,I}\cr}$ transform as a symplectic
vector too. The Schwinger-Zwanziger quantization condition restricts
these charges to a lattice.
Invariance of this lattice restricts the symplectic
transformations to a discrete subgroup $Sp(2m,\Zbar)$.
We will not be concerned with this quantum aspect in this article.

To end this section, we recall some general properties
of symplectic transformations that we will use in the following sections.
For more details and proofs we refer to
appendix~\ref{app:proofssymp}.

If a
$2m\times m$ matrix
\begin{equation}
V=\pmatrix {X^I\cr Y_I}   \label{defmatV}
\end{equation}
 is of rank $m$, there
exist (lemma~\ref{lemma553}) symplectic
transformations $\tilde V={\cal S}V$ such that the upper half of $\tilde V$
constitutes  an invertible $m\times m$ matrix.
For such a $2m\times m$ matrix $V$ with invertible upper part $X$,
one can define a square matrix
\begin{equation}
{\cal N}=Y\,X^{-1} .   \label{NXY}
\end{equation}
If we take all $m$ columns of $V$ to transform as a symplectic vector,
the matrix ${\cal N}$ transforms under symplectic transformations into
\begin{equation}
\tilde{\cal N}=(C+D{\cal N})(A+B{\cal N})^{-1}  \label{tildecalN}
\end{equation}
with ${\cal S}$ as in \eqn{defS}.
The inverse of $A+B{\cal N}$ exists when ${\cal N}$ is a
symmetric matrix with a negative definite imaginary part;
the negative definiteness and symmetry are
preserved by the transformation in \eqn{tildecalN}
(corollary~\ref{gevolg}).

\section{Rigid special geometry}\label{ss:rigid}

\subsection{The vector superfield}
The superfield which is our starting point is the chiral
superfield. The $N=2$ superspace is built with anticommuting
coordinates $\theta^i_\alpha$, and $\theta_{\alpha i}$, where
the position of the index $i$ indicates the chirality\footnote{Recent
reviews on several aspects of $N=2$ supersymmetry are in
\cite{fresoriabook} and \cite{trsummer}. We use
the conventions explained in the appendix of the latter.}. A
{\em chiral} superfield is obtained by imposing the constraint
$ D^{ \alpha i}\Phi=0 $, where
$D^{\alpha i}$ indicates a (chiral) covariant derivative in
superspace. The superfield $\Phi$ is complex, and can be
expanded as
\[ \Phi= A + \theta^i_\alpha  \Psi_i^\alpha +
{\cal C}^{\alpha\beta} \theta^i_\alpha  \theta^j_\beta B_{ij}+
\epsilon_{ij} \theta^i_\alpha  \theta^j_\beta  {\cal F}^{\alpha\beta}
+\ldots \ .\]
One ends up with the set of fields
\begin{equation}
\left( A, \Psi_i, B_{ij}, {\cal F}^{\alpha\beta}, \Lambda_i, C \right) \ .
\end{equation}
Here, $A$ and $C$ are complex scalars, and $\Psi_i$ and
$\Lambda_i$ are two doublet of spinors.
The field  $B_{ij}$ is symmetric.
Finally, as to  ${\cal F}^{\alpha\beta}$,
due to the chirality and the symmetry, it can be written as
${\cal F}^{\alpha\beta}= \sigma^{\alpha\beta} _{ab}{\cal F}^{ab\, -}$, where
${\cal F}^{ab\,-}$ is an antisymmetric antiselfdual tensor.
The complex conjugate superfield $\bar \Phi$ contains the corresponding
selfdual part.

In $N=2$  minimal off-shell multiplets have 8+8 real components. The chiral
multiplet has 16+16 components and is a reducible multiplet. The
{\em vector} multiplet \cite{vectorm} is an irreducible 8+8 part
of this chiral multiplet. The remaining components are combined
into a {\em linear} multiplet, see below. The
reduction is accomplished by an additional constraint, which in
superspace reads
\begin{equation}  \left( \epsilon_{ij}\bar
D^i\sigma_{ab} D^j\right)^2\bar  \Phi=\mp 24\Box  \Phi \ .
\label{constrvector}\end{equation}
Both signs are possible as constraint; we will choose the upper one.
In components, this is equivalent to the conditions
\begin{eqnarray}
H&\equiv& C+2\partial_a\partial^a \bar A =0   \nonumber\\
L_{ij}&\equiv &B_{ij}- \epsilon_{ik}\epsilon_{j\ell }\bar B^{k\ell } =0\nonumber\\
\phi_i&\equiv &\dsl \Psi^i-\epsilon^{ij}\Lambda_j=0\nonumber\\
E^a&\equiv &\partial_b ( {\cal F}^{+\,ab}-{\cal F}^{-\,ab})=0\ , \label{vectorconstr}
\end{eqnarray}
up to some constants%
\footnote{These constants are relevant for
Fayet-Iliopoulos terms \cite{FI,DWVP}, used recently for partial breaking
of $N=2$ supersymmetry \cite{APT}.}.

The equation on $B_{ij}$ is a reality condition. It leaves in
$B_{ij}$ only 3 free real components. Two of the other equations
define $\Lambda$ and $C$ in terms of $\Psi$ and $A$. The remaining
one is a Bianchi identity for ${\cal F}_{ab}$,
expressing that locally it is the derivative of a vector potential:
hence the name vector multiplet for this constrained  multiplet.
To distinguish it from the general chiral superfields, we use a different
notation for the independent fields of the vector multiplet, viz.
$\left( X; \Omega_i;Y_{ij};{\cal F}^-_{ab}\right)$.

\subsection{Actions for $N=2$ vector multiplets}
\label{N=2actie}
A recipe for obtaining supersymmetric actions for these vector multiplets
has been known for some time \cite{PKTN2,DWVP}.
For an arbitrary holomorphic function $F$
of $n$ constrained chiral multiplets $\Phi^A$ ($A=1,\ldots ,n$),
the superfield $F(\Phi^A)$ is again chiral.
Upon integration over the chiral $N=2$ superspace the
action for the vector multiplets follows:
\begin{equation}\int d^4x\int d^4\theta\ F(\Phi^A)\ + c.c.
\label{actsuperspace}\end{equation}
Expanding, this  gives rise to a scalar field Lagrangian
\begin{equation}\label{spin0act}
{\cal L}_0 = -g_{A \bar B}(X,\bar X)\,
\partial_\mu X^A \partial^\mu \bar X^B
\ ,    \label{L0}
\end{equation}
while the coupling of the scalars to the vectors is
described as in \eqn{genL01} by the matrix
\begin{equation}
{\cal N}_{AB}= \bar F_{AB} \equiv \frac{\partial}{\partial\bar X^A}
\frac{\partial}{\partial\bar X^B}\bar F\ , \label{defcNN2}
\end{equation}
where we introduced the notation
\begin{equation}
F_A \equiv \frac{\partial}{\partial X^A} F(X)\ ; \qquad
\bar F_A \equiv \frac{\partial}{\partial \bar X^A} \bar F(\bar X)\ ,
\label{defFA}
\end{equation}
and so on for multiple indices.
The metric in \eqn{L0} is
\begin{eqnarray}
g _{A \bar B}(X,\bar X) & = &\partial_A\partial_{\bar B}K(X,\bar X)
\nonumber\\
K(X,\bar X)&=&i (\bar F_A(\bar X) X^A-F_A(X) \bar X^A)\ ,
\label{metrspc}
\end{eqnarray}
the first equation expressing that it is of \Ka ian type.
\par
As is common for sigma models, the scalars can be interpreted
as coordinates on a manifold,
say ${\cal M}$, at least locally on some chart.
The scalar Lagrangian provides the metric on ${\cal M}$,
thus turning it into a \Ka\ manifold.
Let $z^\alpha$, for $\alpha=1, ..., n$, be complex coordinates in some patch
such that the $X^A$ depend on them holomorphically%
\footnote{This implies the choice of a complex structure.}.
A more general parametrisation of the scalar degrees of freedom is
obtained by using the $z^\alpha$  to describe them.
The \Ka\ potential depends on the
$z^\alpha$ but only through the holomorphic $X(z)$ dependence,
so that the metric retains its hermitian form as in \eqn{L0}.
For definiteness, we give the usual normalisation for
the \Ka\ form:
\begin{equation}
{\cal K}\equiv \frac{i}{2\pi}g_{\alpha\bar \beta}\,dz^\alpha\wedge
d\bar z^\beta=\frac{i}{2\pi} \partial_\alpha \partial_{\bar \beta} K\,
dz^\alpha\wedge d\bar z^\beta=\frac{i}{2\pi} \partial\bar \partial K\ .
\label{defcK}
\end{equation}

Note that the positivity condition for the \Ka\ metric,
\begin{equation} g_{\alpha\bar \beta}= 2\left( \partial_\alpha X^A\right) \ \Im
F_{AB}\ \left( \partial_{\bar \beta}\bar X^B\right)\ , \label{rigmetr}
\end{equation}
is the same as
the requirement of negative definiteness of $\Im{\cal N}_{AB}$, and thus
guarantees the correct sign for the kinetic energies of the vectors.
As already mentioned,
this condition is preserved by symplectic transformations \eqn{tilNN}
(see corollary~\ref{gevolg}).
A further consequence of this positivity condition is that the
following matrix is guaranteed to be invertible:
\begin{equation}
{e_{\alpha}}^A \equiv \partial_\alpha X^A\ .
\end{equation}
One obvious choice of coordinates is then to identify the $z^\alpha$
with $X^A$, which reduces $e_\alpha{}^A$ to the unit matrix. These
coordinates are called special coordinates.
Whereas one can make general coordinate transformations
on the $z^\alpha$, the $X^A$ have a meaning as lowest coordinates of
superfields with constraints which are only invariant under {\em linear}
transformations. That difference, which was stressed in \cite{CdAF},
will play an important role below.

Different functions $F(X)$ may give rise to the same Lagrangian (up to a
total divergence):
\begin{equation}   \label{equivF}
 F\approx F +a+ q_A X^A + c_{AB}X^A X^B \ ,
\end{equation}
where $a$ and $q_A$ can be complex, but $c_{AB}$ should be real.
These transformations are an invariance of the \Ka\ metric
\eqn{metrspc}, and add to ${\cal N}$ a real number, which leads in
\eqn{genL01} only to an extra total
divergence term.

\subsection{Duality transformations in superspace}
\label{ss_dualtf}
There is a remarkable interplay between the symplectic
duality transformations and the supersymmetry.

The action of symplectic transformations on the vector fields,
\eqn{FGsympl}, requires that also the matrix ${\cal N}$
transforms in a specific way, \eqn{tilNN}. The required
transformation  results if one assumes that
\begin{equation}
V = { X^A \choose F_A }\     \label{symplV}
\end{equation}
is a symplectic vector. However, $X^A$ was taken to be the scalar of a
$N=2$ {\em vector} superfield, whereas $F_A$ generically is a scalar
of a {\em chiral} superfield.
To make this situation more
transparent, we adopt a superspace approach\footnote{This way of
presenting the duality transformations came up during a discussion
with P. West.}.

Given a set of chiral superfields and
a holomorphic function $F(X^A)$, the following superfields can be built
\begin{eqnarray}
\Phi_X^A=\Phi^A &=& X^A + \theta^i \Psi_{i}^A + \ldots \ ; \nonumber\\
\Phi_{F,A}=\frac{\partial}{\partial\Phi^A}F(\Phi) &=&
F_A(X) + \theta^i \Psi_{F,Ai} + \ldots \ .
\end{eqnarray}
Here $\Phi^A$ are independent chiral superfields, and the components
of $\Phi_{F,A}$ are functions of the components of $\Phi^A$, e.g.
\begin{equation}
\Psi_{F,Ai}=F_{AB}\Psi_{i}^B\ ;\qquad
{\cal F}^{-\,ab}_{F,A}= F_{AB}{\cal F}^{-\,ab\,B}
-\ft12F_{ABC}\epsilon^{ij}\bar \Psi_{i}^B\sigma^{ab}\Psi_{i}^C\ .
\label{componentsF}\end{equation}
Both sets are combined into a $2n$
component superfield which at the same time should be a symplectic
vector:
\begin{equation}
{ \Phi_X^A \choose \Phi_{F}^A } \equiv \Phi_V = V + \theta^i \Psi_{V,i} + \ldots \ .
\end{equation}
It is therefore natural to impose \eqn{constrvector} not
just on the upper components of $V$, as we did in subsection~\ref{N=2actie}
to reduce the chiral superfields to vector superfields,
but on all components:
\begin{equation}  \left( \epsilon_{ij}\bar
D^i\sigma_{ab} D^j\right)^2\bar  \Phi_V = - 24\Box  \Phi_V \ .\label{constrvectorPhi}
\end{equation}
It turns out that
the lower components of this constraint are the field equations of the
action \eqn{actsuperspace}. So these equations contain {\em all} the
dynamical constraints on the $2n$ chiral superfields.
In particular one finds 
\begin{equation}
\partial_b { {\cal F}^{+\,ab,A} - {\cal F}^{-\,ab,A}
\choose {\cal F}_{F,A}^{+\,ab} - {\cal F}_{F,A}^{-\,ab} } = 0 \ .
\end{equation}
corresponding to $E_a$ in \eqn{vectorconstr}. Inserting the explicit form for
${\cal F}_F^{-\,ab}$, \eqn{componentsF}, this becomes
\begin{equation}
\partial_b { \Im {\cal F}^{+\,ab,A} \choose \Im\left( {\cal N}_{AB}
{\cal F}^{+\,ab,B}+\ft12F_{ABC}\epsilon^{ij}\Omega_{i}^B\sigma^{ab}\Omega_{i}^C
 \right) } = 0 \ .
\end{equation}
We changed notation from $\Psi$ to $\Omega$, since now we are working
with vector multiplets.
The upper equations are the Bianchi identities that reflect this.
The lower components are the field equations for the vector of the
full action; for $\Omega=0$, one recovers those that went with
${\cal L}_1$ of the previous section.

The set of constraints \eqn{constrvectorPhi} is invariant under real
linear transformations $GL(2n,\Rbar)$ on $\Phi_V$.
However, requiring that the
lower superfields in the vector $\Phi_V$ are again derivatives of
some function $F(\Phi)$ imposes the integrability condition
$\partial_{[A} \tilde F_{B]}=0$, which leads again to
\eqn{symplABCD}, restricting the invariances of the constraints to
$Sp(2n,\Rbar)$. This could be expected from the vector part
exhibit in the previous paragraph.

In conclusion, the full superfields $\Phi_V$ are now symplectic vectors,
in particular the lowest components \eqn{symplV}.
The \Ka\ potential can now be written as a symplectic invariant:
\begin{equation}
K= i \langle V,\bar V\rangle \ , \label{Karigidsympl}
\end{equation}
making obvious the symplectic invariance of the scalar field action%
\footnote{Note that when one identifies
the coordinates with the $X^A$ then the symplectic transformation
induces also a coordinate transformation $\tilde X^A=A^A{}_B X^B+B^{AB}F_B(X)$.
This coordinate transformation,
combined with the symplectic transformation,
is only an invariance of the action if
the function $\tilde F(X)$ is equal to $F(X)$, which is only true for
a subset of the symplectic transformations, which are the {\em proper symmetries}
(or hidden symmetries) discussed
e.g. in \cite{DWVP,BEC,prtrquat,hidden}
(within supergravity).\label{fn:scsymplinv} }.

Introducing arbitrary coordinates $z^\alpha$ on ${\cal M}$ as in the
previous section, the gradient of
$V$ defines the following $2n \times n$ matrix~:
\begin{equation}
U_\alpha \equiv {{e^A}_\alpha \choose h_{A\alpha}} \equiv \partial_\alpha V\ .
\label{defUalpha}
\end{equation}
The metric,
\begin{equation}
g_{\alpha\bar \beta}=\partial_\alpha\partial_{\bar \beta}K=i\langle
U_\alpha,\bar U_{\bar \beta}\rangle \ ,    \label{riggUU}
\end{equation}
is assumed to be positive. This implies that $e^A{}_\alpha$ is
invertible (by a simplification of the argument in
lemma~\ref{lem:invbarVU}). Therefore
we can use \eqn{NXY} to define a matrix  which transforms
appropriately under the symplectic transformations.
The matrix ${\cal N}$ introduced in \eqn{defcNN2}
is in fact the complex conjugate of this,
\begin{equation}
{\cal N}_{AB} = \bar h_{A\bar\alpha} {\bar e^{\bar\alpha}}{}_B\ .
\label{defcNrisympl}
\end{equation}
In particular, it will automatically be symmetric.

The \Ka\ potential is invariant under symplectic rotations. But as is
well-known this invariance is not really required. If we combine
different patches so as
to cover a \Ka\ manifold, the following transformation of $K$ preserves
the \Ka\ property relating the local expressions for the metric and the potential:
\begin{equation}
K_{(i)}(z,\bar z)= K_{(j)}(z,\bar z) + f_{ij}(z) + \bar f_{ij}(\bar z)\ ,
\end{equation}
where $f_{ij}$ is a holomorphic function on the overlap region between the
patches. This is usually denoted as a \Ka\  transformation.
In terms of the symplectic vector it translates into the freedom to
perform a constant complex shift, $V' = V + b$, where $b$ is some 2n-component
complex vector.
Another way to explain the origin of the inhomogeneous shift is that it reflects
the linear part in the equivalence relation \eqn{equivF} relating different functions.
The demand that the shift be constant originates in the fact that it
should not affect ${\cal N}$.
Likewise a constant phase shift of $V$ neither changes $K$ nor ${\cal N}$.
In conclusion, the group of symplectic transformations
can be extended with these constant shifts: we will denote%
\footnote{Warning on the notation: the restriction to $\Rbar$
applies to the homogeneous part only. The
inhomogeneous part contains {\em complex} shifts.}
this extension as $ISp(2n, \Rbar)\times U(1)$, where the last factor reflects
the phase rotations.

In this context it is clear that $X^A$ and the $z^\alpha$
have a different role to play, even though the {\em values}
of the $X^A$ can be used as  coordinates ('special coordinates')
as mentioned before.
In the viewpoint we adopt, the $X^A$ are part of $V$ which should be
considered as a section of a specific vector bundle over ${\cal M}$,
whereas $z^\alpha$ are
coordinates in some chart of ${\cal M}$. In the bundle there are two
groups acting {\em independently}. First, there are reparametrisations acting
on the $z^\alpha$, not transforming the sections.
Second, one has the structure group (i.c. $ISp(2n,\Rbar)\times U(1)$)
acting on the fiber.

\subsection{Definitions of rigid special K\"ahler manifolds}

Now we turn to our definitions of a rigid special \Ka\ manifold. The
main criterion is that these manifolds should be obtainable from the scalar
sector of (rigid) supersymmetric models. Historically as well,
it was the analogue of this definition in the supergravity case
to be discussed in the next section,
that was the basis for various other proposed formulations.

In the first definition we use the construction from a
chiral superfield $F$, as sketched above.

\subsubsection{Definition 1}\label{rsg:1}

An $n$-dimensional K\"ahler\footnote{We always suppose here and
below that the metric is positive definite.} manifold is said to be special \Ka\
if it satisfies the following
conditions:
\begin{enumerate}
\item On every chart there are $n$ independent holomorphic functions
$X^A(z)$, where $A=1, \ldots, n$ and
a holomorphic function $F(X)$ such
that
\begin{equation} K(z,\bar z)=i \left( X^A\frac{\partial}{\partial\bar X^A}\bar F(\bar X)-  \bar X^A
\frac{\partial}{\partial X^A} F(X)\right)\ ;
\label{Kapot}
\end{equation}
\item
On overlaps of charts $i$ and $j$ there are transition functions
of the following form:
\begin{equation}
\left( \begin{array}{c}
 X \\ \partial F \end{array}\right)_{(i)} = e^{ic_{ij}} M_{ij}
\left( \begin{array}{c}
 X \\ \partial F\end{array}\right)_{(j)}+ b_{ij}\ , \label{ISpn}
\end{equation}
with
\begin{equation} c_{ij}\in \Rbar\ ;\qquad M_{ij} \in \symp{2n}\ ;\qquad b_{ij} \in
\Cbar^{2n} \ ;\end{equation}
\item The transition functions satisfy the cocycle condition on overlaps
of 3 charts.
\end{enumerate}

{}From this definition it should be clear that the essential
 concept is some specific vector bundle over
a \Ka\  manifold.
To reveal the link with $N=2$ supersymmetric models, first note that the $X^A$ coincide with the
complex scalar fields of the previous section. The holomorphy of the function $F$ is then
necessitated by the demand that $F(X)$ be the lowest component of a chiral superfield
containing the lagrangian. $F$ thus corresponds to the prepotential in the field
theoretic picture. As has been discussed in section~\ref{ss_dualtf}
the transition
functions in \eqn{ISpn} allow  an interpretation as duality transformations.

It is clear that this definition applies to the manifold ${\cal M}$
as constructed from the action in
the previous subsections. Conversely, from the ingredients
of definition~1, one has directly at one's disposal the \Ka\ manifold
for the scalar field action, the matrix ${\cal N}$, defined by
\eqn{defcNN2}, for the spin-1 action, and also all other functions
which enter in the construction of the rest
of the action (as this is defined by \eqn{actsuperspace}).
 \vspace{3mm}

The previous definition heavily relies on the prepotential function
$F(X)$, which is not a symplectically invariant object.
{}From the supersymmetric field theory point of view
it is the most natural definition, though. In a second
definition, inspired by Strominger's definition \cite{special} for
local special geometry, we rest more heavily on
the symplectic bundle structure.

\subsubsection{Definition 2}
\label{rsg:2}
A rigid special \Ka\ manifold is a \Ka\ manifold ${\cal M}$
for which
\begin{enumerate}
\item there exists a  $U(1)\times I\Sp{2n}$ vector bundle over ${\cal M}$
with constant transition functions, in the sense
of \eqn{ISpn}, i.e. with a complex inhomogeneous part. This bundle
should have a holomorphic section $V$ such that the
\Ka\ form \eqn{defcK} is
\begin{equation} {\cal K} = -\frac{1}{2\pi}
\partial\bar{\partial}\langle  V,{\bar V}\rangle \ ,
\label{Kaform}
\end{equation}
\item and such that
\begin{equation}
\label{cond3def2}
\sinprod{\partial_\alpha V}{\partial_{\beta} V} = 0\ .
\end{equation}
\end{enumerate}

The brackets stand for the symplectic inproduct introduced in the previous section.
To establish the equivalence of both formulations one first notes
that the bundle properties~1  from definition~2 are made explicit
in definition~1
both through \eqn{ISpn} and the consistency condition~3.
One further notices that \eqn{Kapot} is the local equivalent of
\eqn{Kaform}, after it has been established that $V$ is of the form
\eqn{symplV}, i.e. that the lower components of $V$ are the derivatives of a
scalar function $F(X)$. Finally, we draw special attention to
\eqn{cond3def2}, which has no analogue in \cite{special}. We
will see that this additional condition is appropriate, since it
leads to the local existence of a function $F$ and guarantees the
symmetry of the matrix ${\cal N}$ which we will define.

We now prove the remaining issue of the local existence of a function
$F$. As argued after \eqn{riggUU},  the matrix $e^A{}_\alpha$
appearing in \eqn{defUalpha}, is invertible.
For the lower components of $V$ one can use the notation
as in \eqn{symplV}, but one should realise that the $F_A$ are
so far just functions of the coordinates $z^\alpha$,
and it remains to be shown that they are of the form \eqn{defFA}.
The invertibility of $e^A{}_\alpha$ allows us to
define $z^\alpha$ as function of $X^A$,
and therefore also the $F_A(z)$ become
functions of $X$. We can then calculate
\begin{equation}
\frac{\partial}{\partial X^A}F_B=e^\alpha{}_A h_{B\alpha}=
e^\alpha{}_A ( h_{C\alpha}e^C{}_\beta )e^\beta{}_B
\end{equation}
and this is symmetric due to \eqn{cond3def2}. This is the integrability
condition that proves the local existence of $F(X)$.

To complete the proof of the equivalence, by constructing
all objects appearing in the action,
we  define ${\cal N}$ with \eqn{defcNrisympl}.
The condition \eqn{cond3def2} then assures that the matrix
${\cal N}$ is symmetric.\vspace{3mm}

We would like to present yet another formulation of rigid special manifolds. As
will be pointed out it will prove to be quite well-adapted to show that
certain subspaces of Riemann surface moduli spaces exhibit special geometry.

\subsubsection{Definition 3}

A rigid special \Ka\ manifold is a complex manifold
${\cal M}$ with on each chart $2n$ closed holomorphic
1-forms $U_\alpha\, dz^\alpha$:
\begin{equation}
\label{rsgdefdeebaru}
\partial_{\bar \alpha}U_{\beta}=0\ ;\qquad
\label{rsgdefdeecommu}
\partial_{\left[\alpha\right.}U_{\left. \beta\right]}= 0\ ,
\end{equation}
such that
\begin{enumerate}
\item
\begin{eqnarray}
\label{rsgdefuu}
\sinprod{U_\alpha}{U_{\beta}}&=& 0 \ ;\\
\label{rsgdefuubar}
g_{\alpha\bar \beta}&=& i \sinprod{U_\alpha}{\bar U_{\bar
\beta}}\qquad
\mbox{is the (positive definite) \Ka\ metric}\ ;
\label{Umetric}
\end{eqnarray}
\item the transition functions on overlap regions are expressed as
\begin{equation}
  \label{rsg3:overgang}
U_{\alpha,(i)}dz_{(i)}^{\alpha} = e^{ic_{ij}} M_{ij} U_{\alpha,(j)}dz_{(j)}
^{\alpha} \ ,
\end{equation}
with $c_{ij}\in \Rbar$ and $M_{ij}\in \symp{2n}$;
\item the cocycle conditions are satisfied on overlaps of 3 charts.
\end{enumerate}

To establish the equivalence with the former definitions  one
should first notice that \eqn{rsgdefdeecommu} imply the local
existence of holomorphic
vectors $V_{(i)}$ such that $U_{\beta,(i)}
= \partial_{\beta}V_{(i)}$. Eq. (\ref{rsg3:overgang}) then guarantees that the $V_{(i)}$ combine
globally into the section $V$ of the product bundle in definition~2. Note the relevance of the
transition functions in \eqn{rsg3:overgang} being constant. With all this
\eqn{cond3def2} coincides with \eqn{rsgdefuu}.
Finally, \eqn{Umetric}) is to be understood as defining the metric in
this approach, which by \eqn{rsgdefdeecommu} is guaranteed to be of
\Ka ian type, but which we should demand to be positive.
In the transition functions, the inhomogeneous term (see \eqn{ISpn})
is no longer present. It reappears in the integration from
$U_\alpha$ to $V$ by \eqn{rsgdefdeecommu}.

\paragraph{Matrix formulation.}

These constraints of this third definition can be concisely formulated in
another way, which will also provide a link to previous formulations
\cite{modssym,f0art,ItalianN2}.
One starts from $n$ symplectic vectors $U_\alpha(z,\bar z)$ over a chart
and their complex conjugates and defines the
$2n \times 2n$ matrix field
\begin{equation}
{\cal V}(z,\bar z)\equiv \pmatrix{U_\alpha ^T \cr \bar U_{\bar \alpha}^T}\ .
\end{equation}
One demands the condition
\begin{equation}
{\cal V} \Omega{\cal V}^T = \pmatrix{0&-ig_{\alpha\bar \beta}\cr
ig_{\beta\bar \alpha} &0}\ ,\label{cVOmcVg}
\end{equation}
where $\Omega$ is the symplectic metric as in \eqn{cSinSP2m}. This is
thus the rewriting of \eqn{rsgdefuu} and \eqn{Umetric}. Then
define
\begin{equation}
\hat {\cal A}_\alpha = \partial_\alpha {\cal V}{\cal V}^{-1}\ ;
\qquad \hat {\cal A}_{\bar\alpha}
= \partial_{\bar\alpha} {\cal V}{\cal V}^{-1}\ ,
\end{equation}
and impose the constraints
\begin{equation} \label{matrixconn}
\hat {\cal A}_\alpha = \left( \begin{array}{cc}  G_{(\alpha,\beta)}{}^\gamma
&C_{(\alpha,\beta)}{}^{\bar \gamma} \\ 0 & 0 \end{array} \right)\ ;\qquad
\hat {\cal A}_{\bar\alpha} = \left( \begin{array}{cc} 0 & 0 \\
\bar C_{\bar\alpha,\bar \beta}{}^\gamma &
G_{\bar\alpha,\bar \beta}{}^{\bar \gamma} \end{array} \right)\ ,
\end{equation}
satisfying the indicated symmetry requirements (the form of
$\hat {\cal A}_{\bar\alpha}$ follows by complex conjugation).
These requirements are equivalent to the conditions \eqn{rsgdefdeecommu}.

One can further determine and simplify the matrix relations by using
\eqn{cVOmcVg}:
\begin{eqnarray}
\left( \hat {\cal A}\right)_\beta{}^\gamma \pmatrix{0&-ig_{\gamma\bar\delta}\cr
ig_{\delta\bar \gamma} &0}&=& \left( \partial_\alpha{\cal V} \Omega{\cal V}^T\right)
_{\beta\delta}\nonumber\\
&=& \partial_\alpha  \pmatrix{0&-ig_{\beta\bar\delta}\cr
ig_{\delta\bar\beta} &0}  - \left( {\cal V} \Omega\partial_\alpha{\cal V}^T\right)
_{\beta\delta}        \label{eqnAG}
\end{eqnarray}
The last term of the second line is the transpose of the left hand
side of the first line. Furthermore, from \eqn{matrixconn} we have
that
\begin{equation}
\left( \hat {\cal A}\right)_\beta{}^\gamma \pmatrix{0&-ig_{\gamma\bar\delta}\cr
ig_{\delta\bar \gamma} &0}=
\pmatrix{iC_{(\alpha,\beta)\delta}&-iG_{(\alpha,\beta)\bar \delta}\cr 0 &0}\ ,
\end{equation}
(indices lowered by the metric $g_{\alpha\bar \beta}$),
i.e. symmetric in $(\alpha\beta)$.  Therefore, taking the second line
of \eqn{eqnAG} minus the transpose of the first, we obtain
\begin{equation}
\pmatrix{iC_{(\alpha,\beta)\delta}-iC_{(\alpha,\delta)\beta}&
-iG_{(\alpha,\beta)\bar \delta}\cr iG_{(\alpha,\delta)\bar\beta }  &0}=
\partial_\alpha  \pmatrix{0&-ig_{\beta\bar\delta}\cr
ig_{\delta\bar\beta} &0} \ .
\end{equation}
It implies first that $C$ is a 3-index symmetric tensor\footnote{One
can check that
$C_{\alpha\beta\gamma}=i\, e^A_\alpha e^B_\beta e^C_\gamma\, F_{ABC}$.}.
Furthermore it implies that $\partial_{[\alpha}g_{\beta]\bar
\gamma}=0$, i.e. that this is a \Ka\ metric. Therefore it is
appropriate to define covariant derivatives with Levi-Civita connection:
\begin{equation}
\Gamma_{\alpha \beta }^\gamma  = g^{\gamma \bar \delta }
\partial_\beta  g_{\alpha \bar \delta }\ ;\qquad
{\cal D}_\alpha U_\beta=\partial_\alpha U_\beta
-\Gamma_{\alpha\beta}^\gamma U_\gamma \ ;\qquad
{\cal D}_{\bar \alpha} U_\beta=\partial_{\bar \alpha} U_\beta\ .
\label{LeviCivita}
\end{equation}
The tensors $G$ are thus these connections, and the differential
equations can be simplified to~\cite{modssym}~:
\begin{eqnarray}
{\cal D}_\alpha  {\cal V}={\cal A}_\alpha {\cal V}  \ ;\qquad
{\cal D}_{\bar \alpha}  {\cal V}={\cal A}_{\bar \alpha} {\cal
V}\nonumber\\
 {\cal A}_\alpha =\pmatrix{0& C_{\alpha \beta}{}^{\bar \gamma } \cr 0&0\cr}  \ ;\qquad
 {\cal A}_{\bar \alpha} =\pmatrix{0&0\cr \bar C_{\bar \alpha \bar \beta}{}^\gamma
&0\cr}  \ .
 \label{rsgmatrx}
\end{eqnarray}
for a symmetric $C_{\alpha \beta \gamma }$.
{}From the commutator of covariant derivatives,
 the following curvature formula is then easily derived
\begin{equation}
R_{\alpha \bar \beta \gamma \bar \delta}=- C_{\alpha \gamma\epsilon}
\bar C_{\bar \beta \bar \delta \bar\epsilon} g^{\epsilon\bar
\epsilon}\ .   \label{Rrigid}
\end{equation}

\subsection{Riemann surfaces moduli spaces}   \label{ss:RS}
After the original formulation of {\em local} special geometry \cite{DWVP},
a notion to be developed in the next
sections, it was soon found out that the defining constraints are realised
in Calabi-Yau moduli spaces \cite{FerStroCand}.
In the study of {\em rigid} $N=2$ super Yang-Mills models,
Riemann surface moduli spaces turn out
to be a candidate to take over the role of Calabi-Yau moduli spaces
\cite{SeiWit}.
A relevant question therefore is the following.
Do there exist
$r$-dimensional subspaces of the full moduli space of genus $g$ Riemann
surfaces that obey the constraints
dictated by rigid special geometry?
A partial answer to this is the main subject of this section.
Definition~3 is the formulation which
contains the data by which such an identification is most easily
established. Most of its constraints are automatically satisfied,
and we will see that \eqn{rsgdefdeecommu} is the crucial one to
verify.
\subsubsection{Useful concepts of Riemann surfaces}
Let us first recall some facts from Riemann surface
theory. For definiteness consider a genus $g$ surface. As is
well-known the first homology group then has $2g$ generators $A^A$,
$B_A$ with index $A=1, \ldots, g$ which can
be fixed to be canonical, i.e. with a particular intersection matrix defined by
the following relations:
\begin{eqnarray}
&&A^A \cap A^B = 0\ ; \qquad \quad B_A \cap B_B = 0\ ;
\nonumber\\ &&A^A \cap B_B = -B_B \cap A^A =
\delta^A_B \ . \label{intersections}
\end{eqnarray}
The $2g$ component vector consisting of integrals of a one-form $\omega$
along a canonical homology basis
\[
{ \int _{A^A} \omega \choose \int _{B_A} \omega }
\]
is referred to as the period vector of $\omega$ w.r.t that homology basis.

For any pair of closed one-forms $\omega$ and $\chi$,
\begin{equation}
\sum_{A}\left[\int_{A^A} \omega \cdot \int_{B_A} \chi -
\int_{B_A} \omega \cdot \int_{A^A} \chi  \right]=\int\int  \omega
\wedge \chi  \ ,    \label{integralformula}
\end{equation}
where the integral at the right hand side is over the Riemann
surface.

Let  $\{\lambda_\alpha\}$ with $\alpha = 1 \ldots g$,
be some basis for the one forms that are holomorphic
w.r.t. the complex structure on the Riemann surface.
The matrices
\begin{equation}
{e^A}_\alpha \equiv  \int_{A^A}\lambda_\alpha \ ;\qquad  h_{A\alpha } \equiv
\int_{B_A}\lambda_\alpha\ ,
\end{equation}
are both invertible.
{}From these one defines the so-called period matrix $\Pi$:
\begin{equation}
\Pi_{AB} = h_{A\alpha} {(e^{-1})^\alpha}_B\ . \label{defperiodm}
\end{equation}

Using \eqn{integralformula} with two holomorphic forms, $\lambda_\alpha$ and
$\lambda_{\beta}$, the right hand side is zero, which is referred to
as Riemann's first relation \cite{grifharr}.
Therefore
\begin{equation}
0 =  \left[e^A_\alpha   h_{A \beta}
-h_{A\alpha} e^A_{ \beta}\right]
=e^A_\alpha \left[
 h_{A \gamma} (e^{-1})^{ \gamma} _B  - (e^{-1})_A^\gamma h_{B\gamma}
\right] e^B_{ \beta} =e^A_\alpha \left[
 \Pi_{AB} - \Pi_{BA}\right]  e^B_{ \beta} \ ,\label{riemann1}
\end{equation}
so that Riemann's first relation expresses the symmetry of the period
matrix.

Using \eqn{integralformula} with the forms $\lambda_\alpha$ and their
complex conjugates $\bar \lambda_{\bar \beta}$, the right hand side
is a positive matrix times $dx\,d\bar x$ (with $x$ the holomorphic
coordinate on the Riemann surface), and the surface can be
oriented such that this is $-i$ times a positive quantity. Therefore
we have
\begin{equation}
0 < i \left[e^A_\alpha  \bar h_{A\bar \beta}
-h_{A\alpha}\bar e^A_{\bar \beta}\right]
=e^A_\alpha \left[i
\bar h_{A\bar \gamma}(\bar e^{-1})^{\bar \gamma} _B
-i (e^{-1})_A^\gamma h_{B\gamma}
\right]\bar e^B_{\bar \beta} =e^A_\alpha \left[
i\bar \Pi_{AB} -i \Pi_{BA}\right] \bar e^B_{\bar \beta} \ ,
\label{riemann2a}
\end{equation}
which is Riemann's second relation \cite{grifharr}~:
\begin{equation} \label{riemann2}
\Im \Pi > 0 \ .
\end{equation}

 It is to be noticed that both Riemann's relations \eqn{riemann1}
and \eqn{riemann2} are invariant under a symplectic change of homology basis,
taking a canonical
basis into another one, preserving the intersection matrix.
This follows from a general fact about symplectic transformations
(see Corollary~\ref{gevolg}).

\subsubsection{Special moduli spaces of Riemann surfaces}
Now we turn to the issue of special geometry in some moduli
space of Riemann surfaces.
It is clear that the properties of the period matrix $\Pi$ make it
extremely well suited to play the role of the matrix ${\cal N}$.
In general however the dimensions do not agree: the period matrix has
dimension $g$, whereas the dimension of the moduli space is generically
larger. We will therefore consider an $n$-dimensional {\em subspace}
of the moduli space. It will turn out that in the simplest case $n=g$
a special geometry structure is almost automatic,
which motivated the consideration of these spaces for the construction
of special geometries in the first place, but we will not limit
ourselves to that case.

Let ${\cal W}$ represent some family of genus $g$
Riemann surfaces parametrized by $n$ complex moduli, where $n\leq g$.
Let
$\gamma_{\alpha },\ \alpha =1, \ldots, n$
be a set of cohomologically independent 1-forms that are holomorphic
with respect to the complex
structure on the Riemann surfaces. For $n<g$ these span
a subspace of the holomorphic 1-forms, $H^{(1,0)}({\cal W})$.
The number of 1-forms chosen is equal to the
complex dimension of the considered moduli space.

Let $A^{A}, \, B_{A}$ with $A=1,\ldots, n$ be some set of
 1-cycles  satisfying again \eqn{intersections}, spanning
 a proper subspace  of the first homology group if $n<g$.
Identify now
\begin{equation}
U_\alpha=\pmatrix{\int_{A^A} \,
\gamma_{\alpha}\cr \int_{B_B}\gamma_\alpha\cr}\ .  \label{link}
\end{equation}
These 'periods' are then complex functions of the moduli, which are
to be identified with the  functions $U_\alpha$ introduced abstractly in
definition~3.

The first constraint of \eqn{rsgdefdeecommu} says that
the moduli should be chosen such that these periods depend
holomorphically on the moduli $z^\alpha$.
A convenient way to realize this automatically is to define the family of
Riemann surfaces in a complex weighted projective space by an
equation that is holomorphic in the variables of the
projective space as well as in the moduli. The 1-forms can then be defined
using the Griffiths residue map \cite{griffiths}. In that case this
holomorphicity condition of $U_\alpha$ is immediate.
We defer the discussion of the second condition of \eqn{rsgdefdeecommu},
and first take a look at the other elements of definition~3.

If $n=g$ then \eqn{rsgdefuu} is \eqn{riemann1}, Riemann's first
relation. Then also \eqn{Umetric}, defining the metric, leads
automatically to a positive definite metric, as it is
\eqn{riemann2a}, Riemann's second relation.
The symplectic
transformations, which appear in \eqn{ISpn}, and
which in field theory embody duality
transformations are now realised by
canonical homology basis changes. Note that the restriction to
integer symplectic matrices,
$Sp(2n,\Zbar )$, which occurs in the quantum theory, is also natural
from this point of view.
Finally, as already mentioned,
$\bar{\cal N}_{AB}$ is the period matrix $\Pi_{AB}$, \eqn{defperiodm}.
If $n<g$, we can not apply \eqn{integralformula}
but we will see some examples in the next subsection that
show how Riemann's relations, combined with symmetries,
may still suffice to prove \eqn{rsgdefuu} and \eqn{Umetric}.
In the general case however, to which we now return, an explicit check
of these conditions seems required. The symplectic group can still be
realised by canonical changes of basis for the subspace of cycles one is
considering, but the restriction of the  period matrix $\Pi_{AB}$
to the subset of forms and cycles, which should play the role of
$\bar{\cal N}_{AB}$, may not automatically have a symmetric and positive
definite imaginary part. Again, in some examples, selecting the
subspaces by some symmetry requirement suffices to remedy this.

Crucial for the metric to be \Ka ian as well as for the existence of
the prepotential $F(X)$ is the second constraint in
\eqn{rsgdefdeecommu}. It is not known what
the geometric significance is, of moduli spaces which satisfy such a
constraint. If the family of Riemann surfaces is defined in an
embedding space (e.g. the weighted projective space mentioned
above), and the cycles can be constructed in a way that is
moduli independent (in some local patch of moduli space),
then the condition translates into an integrability
condition on the 1-forms:
\begin{equation} \label{intg-1forms}
\partial_{[\alpha}\gamma_{\beta]} = (d\eta)_{\alpha\beta}\ .
\end{equation}
If the right hand side is zero, as in most examples we know, then
locally there should exist a (meromorphic) 1-form $\lambda$
whose derivatives give
the holomorphic 1-forms. The form
$\lambda$ should have zero residues at its poles, such that its integrals
over the 1-cycles, which are used to construct the symplectic vector $V$,
are well-defined.
Notice that we have used a coordinate-index $\alpha$
to label the $n$ one forms $\gamma_\alpha$, thus implicitly associating
each of the one forms to a modulus.
Apart from matching the dimensions one has to make sure that
the \eqn{intg-1forms} actually holds, for the chosen moduli
subspace to be rigid special \Ka.

The conditions on the transition functions and cocycle conditions
are also  automatically satisfied if the cycles and forms are either
a complete basis of those available on the Riemann surface, or if
they are selected by some symmetry requirement.

At present it is not clear whether the
outlined procedure may yield any rigid special \Ka\ subspace in the generic
case for arbitrary $g$. This
has to be contrasted with the case of Calabi-Yau moduli spaces, which always
obey the local special geometry constraints. This will be discussed later.
\subsubsection{Examples}
As a first example, where $n=g$, we  sketch briefly how
the proposed $SU(n+1)$
curves \cite{klemmlerche} in
$N=2$ SYM fit into the scheme. The starting point is some genus $n$
family of Riemann
surfaces, corresponding to the rank of the gauge group.
In general the full moduli space is
$3g-3$ (complex) dimensional, but it is reduced from the outset to
$2g-1$, taking only the class of hyperelliptic surfaces into consideration. A further
reduction a posteriori, giving rise to the right number of moduli (i.e.
$n$ or equivalently,
$g$) is performed through symmetry arguments (for a detailed discussion
see
\cite{klemmlerche}). One thus ends up with a $g$ moduli dependent genus $g$ family. The
remaining integrability condition \eqn{intg-1forms} is met as an explicit
expression for the meromorphic 1-form $\lambda$ is derived. As a consequence, the reduced
moduli space is rigid special \Ka.

It is to be noticed how the notion of the symplectic group of Definition~3 arises in this
context. First taking into account the identification \eqn{link},
the period matrix $\Pi$
of the Riemann surfaces is the kinetic vector coupling matrix $\bar{\cal N}$.
With this the symplectic transformations in the context of rigid special \Ka\ manifolds can be
thought of as the canonical homology basis changes.
Apart from the natural interpretation of
the structure group ($ISp(2g,\Rbar)$ up to phase transformations) in the context of surfaces,
the key point in matching the genus $g$ and
$n$, the dimension of the associated moduli space,  resides in that the homology and
cohomology groups have the right dimensionality to insure the validity of both Riemann's
relations \eqn{riemann1},\eqn{riemann2}.

In the above the main step was the isolation of the special \Ka\ moduli
subspace within the full $3g-3$ dimensional moduli space of genus
$g$ curves.\vspace{3mm}

An extra complication arises when $n<g$, which we find e.g. in
\cite{sundborg}. Specific families of surfaces are presented there
as candidate curves for solving
$N=2$ SYM with gauge groups $SO(2n+1)$ and $SO(2n)$ respectively.
One considers in both cases Riemann surfaces of genus $g=2n-1$,
represented by hyperelliptic curves whose defining equation
$y^2=P(x,z^\alpha)$ has a symmetry $P(-x,z^\alpha)=P(x,z^\alpha)$.
There is a meromorphic form $\lambda$ which is odd under the
symmetry, and whose derivatives with repect to the moduli give $n$ (odd)
holomorphic 1-forms. Therefore the conditions \eqn{rsgdefdeecommu}
are already satisfied. The $2g=2(2n-1)$ one-cycles can be split in
$2(n-1)$ even and $2n$ odd under $x\leftrightarrow -x$. So one can
restrict to the latter for the setup described above. The forms
$\gamma_\alpha=\partial_\alpha \lambda$ can be completed to a basis of
the $g$ dimensional cohomology of $(1,0)$-forms by $(n-1)$ forms which
are even under the symmetry (all this was done explicitly in
\cite{stefanlic}). Therefore, e.g. the full matrix of integrals
of $(1,0)$ forms over the $g$ $A$-cycles, is block diagonal, containing
$e^A_\alpha$ as one of its blocks. The fact that this full matrix is
invertible then implies the invertibility of $e^A_\alpha$. The same
block diagonal structure appears everywhere in the matrices which enter
in Riemann's two identities, which therefore still imply \eqn{rsgdefuu}
and the positivity of the metric in \eqn{Umetric}.

Also for other associations of the quantum theory of rigid $N=2$
supersymmetry - Yang-Mills models with moduli spaces of Riemann surfaces
\cite{rigidSUSYYM} similar procedures can be followed to prove that
they satisfy the conditions of definition~3.

\section{(Local) Special geometry}\label{ss:local}
\subsection{Supergravity actions}\label{sugract}
We have seen that we have at our disposal quite a few equivalent
formulations of rigid special geometry.
In the remaining sections we would like to find out in how far these results are transferable to
the supergravity or Calabi-Yau case.

The name special geometry \cite{special} has been given to the manifold determined by
the scalars of vector multiplets in $N=2$ supergravity. The first
construction~\cite{DWVP} of these actions relied on superconformal tensor calculus,
the principle of which is that one starts with defining multiplets of the
superconformal group, and then at the very end breaks the symmetry down to the super
Poincar\'e group\footnote{For reviews, see e.g. \cite{revN2}, or
\cite{LondonN2sg} for a shorter recent one.}.
This, at  first sight cumbersome, procedure in fact
turns out to simplify
life. The Poincar\'e supersymmetric action contains many terms of which
the origin is not clear, unless one recognises how they arose naturally in
the superconformal setup.

For $d=4,N=2$ the superconformal group is
\begin{equation}
SU(2,2|N=2) \supset SU(2,2)\otimes U(1)\otimes SU(2)\ .\label{scgN2}
\end{equation}
In the bosonic subgroup displayed, the conformal group is identified as the
$SU(2,2)=SO(4,2)$ factor. On the fermionic side there are two
supersymmetries $Q^{i}$ and two special supersymmetries $S^{i}$.
The multiplet gauging the superconformal group is called the Weyl
multiplet, and has as independent fields
\begin{equation}
\{ e_\mu^a,\,b_\mu,\, \psi_\mu^i,\,A_\mu,\,{\cal V}_\mu{}^i{}_j,\,
T^-_{ab},\, \chi^i,\, D\}\ .
\end{equation}
These are the gauge fields for general coordinate transformations,
dilatations, $Q^{i}$, $U(1)$ and $SU(2)$. The extra antiselfdual
antisymmetric tensor, a doublet of fermions and a real scalar are
included to close the algebra.
In order
to gauge fix the superfluous symmetries some vector multiplet and a second
compensating multiplet (e.g. a hypermultiplet) are introduced.
After the breakdown to the super Poincar\'e group, of the Weyl multiplet only the
vierbein and the gravitinos will remain as physical fields. Likewise,
the hypermultiplet  disappears from the action by
gauge fixing the $SU(2)$ and by the field equations of $D$ and $\chi^{i}$.

For a theory with $n$ physical vector multiplets we introduce $n+1$ vector
multiplets to start with. One of these is to be identified as a
compensating multiplet, of which the scalar disappears upon fixing the
dilatational and $U(1)$ gauge, the fermion by the gauge fixing of
$S$-supersymmetry, while its vector becomes the physical graviphoton.
In the end one realises that $n$ complex scalars, $n$
doublets of spinors, and $n+1$ vectors constitute the physical content
originating from the vector multiplets.

These vector multiplets consist of  the following fields:
\newcommand{\ghe}[3]{\stackrel{\textstyle #1}{\scriptstyle (#2,#3)}}
\begin{equation}
\stackrel {\textstyle(}{\phantom{\scriptstyle (}}
\ghe{X^I_{\phantom{i}}}1{-1},
\ghe{\Omega_i^I}{\frac 32}{-\frac 12},
\ghe{{\cal A}_\mu^I}00, \ghe{Y_{ij}^I}20
\stackrel {\textstyle)}{\phantom{\scriptstyle (}}\qquad
\stackrel{\textstyle\mbox{with}}{\phantom{\scriptstyle (}}\qquad
\stackrel{\textstyle I=0,1,...,n.}{\phantom{\scriptstyle (}}
\label{vectormult}\end{equation}
We have indicated for each component field
the weights  $(w,c)$
defining their transformation under the dilatation and $U(1)$
transformations in the superconformal group:
$\delta\phi=w\,\phi\,\Lambda_D +i\,c\,\phi\,\Lambda_{U(1)}$.

The action is then built exactly as in the rigid case,
i.e. through the chiral superspace integral of a holomorphic function
$F(X)$, called a {\em prepotential}. The scaling symmetry now imposes
an extra requirement: $F(X)$ must be
{\em homogeneous of degree two} in $X^I$,
where $I=0,1,...,n$.

The $X^I$ span a ($n+1$)-dimensional
complex space, but as a result of the dilatation and $U(1)$
symmetry, the  physical scalars parametrize an $n$-dimensional
complex hypersurface which will turn out to be \Ka.
We now proceed to fix these gauges, while building in the freedom to
perform \Ka\ transformations on that hypersurface.

The gauge fixing can be performed in a way such that the symplectic
structure remains manifest, and an arbitrariness in the gauge fixing
of the phase transformations of $X^I$ gives rise to the \Ka\
transformations. To perform such a gauge fixing,
it is convenient to write the $n+1$ complex fields $X^I$ in terms of
$n+2$ complex fields $a$ and  $Z^{I}$:
\begin{equation}
X^I= a Z^I\ ,   \label{X2Z}
\end{equation}
which implies the extra gauge invariance
\begin{equation}
a'= a\, e^{\Lambda_{aZ}}\ ; \qquad Z'^I=Z^I\,e^{-\Lambda_{aZ}} \ ,
\label{pre-Kahlertransf}
\end{equation}
where $\Lambda_{aZ}$ is an arbitrary complex gauge parameter.
Since all $X^I$ transform with the same weights under dilatations and $U(1)$
transformations, this allows us to separate out the action of these
transformations by letting them  act on $a$ with the same weights,
so that $Z^I$ is invariant.
Under the combined action of these gauge
transformations  and dilatations and  $U(1)$ transformations we have then
\begin{eqnarray}
 X'^{I} &=& e^{\Lambda_{D}-i\Lambda_{U(1)}} X^{I}\ ;\\
 a' &=& e^{\Lambda_{D}-i\Lambda_{U(1)}+\Lambda_{aZ}} a\ ; \\
 Z'^{I}&=& e^{-\Lambda_{aZ}}Z^{I}\ .
\end{eqnarray}
Now we go on to fix these gauges.

In the action, the curvature is coupled to the scalars via the term
\begin{equation}
i(\bar X^I F_I - \bar F_I X^I)e\ R\ .
\end{equation}
A standard Einstein action can be obtained by choosing as dilatational
gauge fixing
\begin{equation} \label{constraint}
i(\bar X^I F_I -\bar F_I X^I)=i\langle \bar V , V\rangle =1
\qquad\mbox{with}\qquad
 V\equiv\pmatrix{X^I\cr F_I} \ .
\end{equation}
We define $K(Z,\bar Z)$ and $N_{IJ}$ through
\begin{equation}
e^{-K(Z,\bar Z)}
\equiv i\left( \bar Z^I F_I(Z)-\bar F_I(Z) Z^I\right)
=-\bar Z^I N_{IJ} Z^J\ ;\qquad N_{IJ}\equiv 2\,\Im F_{IJ} \ .
\label{Kahlerpot}
\end{equation}
The homogeneity of $F$ implies that $F_I(X)=aF_I(Z)$ and $N_{IJ}(X)=N_{IJ}(Z)$).
The dilatational gauge fixing \eqn{constraint} then amounts to setting
\begin{equation}
|a|^2=e^{K(Z,\bar Z)}\ .  \label{a2K}
\end{equation}

The $U(1)$ gauge invariance of the superconformal group
is fixed by specifying the phase of $a$: we take $a$ to be real and positive.

The remaining unfixed gauge invariance is now the transformation
\eqn{pre-Kahlertransf}, accompanied by a $U(1)$ transformation, with parameter
\begin{equation}
2i \Lambda_{U(1)}= \Lambda_{aZ} -\bar \Lambda_{aZ}\ .
\label{decompU1}
\end{equation}
This gauge freedom can be fixed by specifying a single (complex) gauge
condition. Generically, this could be done by specifying a
rather arbitrary condition $F(Z^I,\bar Z^I)=0$.
Here, however, we impose that the gauge condition be
an (inhomogeneous) equation that is {\em holomorphic} in  $Z^I$.
We do not need to make an explicit choice for this equation at this point.
However, we have to discuss how to relate two such choices (both
holomorphic) with each other. The relation is
\begin{equation}
Z'^I=Z^I\,e^{-\Lambda_K(Z)} \ .
\label{Kahlertransf}
\end{equation}
Note that this transformation, in accordance with \eqn{decompU1},
implies a $U(1)$ rotation with parameter
\begin{equation}
2i \Lambda_{U(1)}= \Lambda_K(Z) -\bar \Lambda_K(\bar Z) \ ,
\label{decompU1K}
\end{equation}
so that the
total effect of this change of gauge is
\begin{eqnarray}
 X'^{I} &=& e^{-\frac{1}{2}(\Lambda_K(Z) -\bar \Lambda_K(\bar Z))} X^{I}\ ;\\
 a' &=& e^{\frac{1}{2}(\Lambda_K(Z) +\bar \Lambda_K(\bar Z))} a\ ; \\
 Z'^{I}&=& e^{-\Lambda_K(Z)}Z^{I}\ .
\end{eqnarray}
The corresponding transformation of $K$ in \eqn{Kahlerpot}  is
\begin{equation}
K'(Z,\bar Z)= K(Z,\bar Z)+\Lambda_K(Z) + \bar \Lambda_K(\bar Z)\ .
\label{K-trans on K}
\end{equation}
This transformation will presently acquire the meaning of a \Ka\ transformation.

The action for the scalars is
\begin{equation}
{\cal L}_0=-e\,N_{IJ}{\cal D}_\mu X^I\,{\cal D}^\mu \bar X^J
\qquad\mbox{with}\qquad
{\cal D}_\mu X^I =(\partial_\mu+iA_\mu)X^I\ , \label{L0sc}
\end{equation}
with $X^I$ given in \eqn{X2Z} and $a$ by the
positive square root of \eqn{a2K}.
The auxiliary gauge field of the $U(1)$ transformation
is eliminated by its field equation:
\begin{equation}
A_\mu=\ft i2 N_{IJ}
\left( X^I\partial_\mu\bar X^J-\partial_\mu X^I\bar X^J\right) \ .
\label{Amusoln}
\end{equation}
The spin-1 action is given by equation \eqn{genL01} with
\beq                  \label{NvanF}
 \N_{IJ}(Z)= \bar{F}_{IJ}(Z)+2i\, \frac{\IM F_{IK}(Z)\: \IM F_{JL}(Z)\: \ZK\,\ZL}
          {\IM F_{KL}(Z)\:\ZK\ZL} \ .
\eeq
The full action has been constructed in \cite{dWLVP}.

The scalar action describes a \Ka\ manifold with coordinates%
\footnote{There are $n+1$ quantities $Z^I$, but after
properly fixing the gauge only $n$ of them are independent: these are
good coordinates if the gauge fixing condition on the $Z^I$ has been chosen
properly.}
$Z^I$ and with \Ka\ potential
\begin{equation}
K(Z,\bar Z)= -\log\left[i
\bar Z^I \frac{\partial}{\partial Z^I} F(Z) -i
Z^I\frac{\partial}{\partial\bar Z^I}\bar F(\bar Z)
 \right]\ .
\end{equation}
The transformation of coordinates $Z^I$ in \eqn{Kahlertransf} induces
the correct change of \Ka\ potential, \eqn{K-trans on K}.
It is clear however that \eqn{Kahlertransf}
is not the most general holomorphic change of coordinates that one
may consider.
Now we regard the $Z^I$ ($I=0,\ldots n$) as holomorphic
functions of the 'physical scalars'
$z^{\alpha}$ ($\alpha=1,\ldots, n$), which constitute a more general
holomorphic coordinate system. As in the case of rigid supersymmetry these
scalars are coordinates on (one patch of) a complex manifold.
By restricing the $Z^I$ to holomorphic functions of the coordinates $z$,
the kinetic term of the scalars maintains its hermitian form.

Having adopted general holomorphic coordinates $z$, we can now consider
arbitrary holomorphic coordinate changes, and the induced \Ka\
transformation. By letting $Z^I$ transform as in \eqn{Kahlertransf}
(but now with the parameter $\Lambda_K$ depending on $z$), it becomes a
section of a fibre bundle\footnote{To be precise, we also allow the symplectic
transformations discussed in the next paragraph.}.
The complete {\em scalar manifold} can be viewed as
a number of patches glued  together, where in each patch the description above
applies. In going from one patch to
another  the $Z^I$ transform under a K\"ahler transformation.
Note that with this interpretation, through the accompanying $U(1)$
transformation \eqn{decompU1K}, the superfield components transform
also. Therefore, they will have to be interpreted as sections of the
appropriate bundles too.

In addition to \Ka\ transformations,
there is another invariance in the theory, viz.
the symplectic duality transformations. As in the rigid case,
both classes of transformations  preserve the K\"ahler metric, and
the system of equations of motion and Bianchi identities for the vectors.
Apart from one exception below, we will only have to consider
the bosonic sector of the theory, but the terms with fermions
have been shown to be invariant as well \cite{DWVP,f0art}.

There is an extra requirement that we want to impose, viz.
the positivity of the kinetic energies. Let us start with the spin-2 sector. We can
only make the dilatational gauge fixing
 \eqn{constraint} provided the left-hand side is
positive from the start. This requirement restricts the domain where a particular
function $F$ can be used to construct an action. See \cite{DWVP} for an example.
The spin-0 sector provides a second condition: the positivity of the metric of the
scalar manifold. We take the positivity of the metric as part of the definition of
a K\"ahler manifold. A third condition might be expected from the spin 1 sector,
but it was proved in \cite{BEC} that the vector kinetic energy is
automatically positive if the other two conditions are satisfied.

There is one important global aspect, which as far as we know is the
only instance where the fermion sector comes in. The fermions impose a quantisation
condition on the \Ka\ form. For a compact gauge
group, the cohomology class of the 2-form gauge field strength is
quantised: if fields transform as $\psi\rightarrow U\psi$, then the
gauge field $A_\mu$, normalised so that it transforms as
$\partial_\mu + i A_\mu \rightarrow U^{-1} (\partial_\mu + i A_\mu) U$,
has a field strength
$F_{\mu\nu}=\partial_\mu A_\nu -\partial_\nu A_\mu$, of which the integral
over an arbitrary 2-cycle is
\begin{equation}
[F]\equiv \int F_{\mu\nu}\,dx^\mu\,dx^\nu = 2\pi in\ ;\qquad n\in \Zbar\ .
\end{equation}
This is analogous to the quantisation of the magnetic charge and is
in a similar setting  nicely explained in \cite{BaggerWittenN1}.
In a more mathematical language this means that the transformation
functions should define a complex $U(1)$ line bundle, whose first
Chern class should be of integer cohomology \cite{Chern}.
If several line bundles  exist, having different $U(1)$ charges
that are multiples of a basic unit, then the most stringent condition
is given by a bundle having this unit charge.

We now apply this consideration to the \Ka\ transformations\footnote{Note that
in this case the base manifold is the scalar manifold instead of spacetime.}.
By \eqn{decompU1K} they are accompanied with a $U(1)$ transformation on
the fields $X^I$, and therefore of the whole supermultiplet.
The fields with the lowest $U(1)$ weight in this multiplet are the fermions,
see \eqn{vectormult}.
Using \eqn{decompU1K}, we get
\begin{equation}
\Omega_i\rightarrow
e^{-\ft14\left( \Lambda_K(z) -\bar \Lambda_K(\bar z)\right)}\Omega_i
\ .
\end{equation}
This implies that the $U(1)$ connection in this bundle is normalised
as\footnote{The connection is
$(-i/2)$ times the gauge field of the superconformal $U(1)$, $A_\mu$, as
can be seen by the covariant derivatives in \eqn{L0sc}, which are written
there for fields of chiral weight which is the double of that of the
fermions.}
\begin{equation}
A_\alpha=-\ft{1}{4}\partial_\alpha K\ ;\qquad
A_{\bar \alpha}=\ft{1}{4}\partial_{\bar \alpha}K\ ,  \label{defAKa}
\end{equation}
and the curvature,
\begin{equation}
F_{\alpha\bar \beta}=-F_{\bar \beta\alpha}
=\ft{1}{2}g_{\alpha\bar \beta}\ ,
\end{equation}
is proportional to the \Ka\ 2-form \eqn{defcK}.
The bundle's first Chern class, which is of integer cohomology,
thus equals $\frac{1}{2}$ times the
de Rham cohomology class of the K\"ahler form:
\begin{equation}
c_1=\ft{1}{2}[{\cal K}]\in \Zbar\ .
\end{equation}
We find that the K\"ahler form should be of even integer cohomology.
If there had been no fermions present in the theory, the same argument applied just to the
bosonic fields would have allowed an arbitrary integer cohomology. In
the mathematical literature \cite{Chern,Wells} K\"ahler manifolds of which the
K\"ahler form is of integer cohomology
are called {\em K\"ahler manifolds of restricted
type} or {\em Hodge manifolds}. With a slight abuse of language we will call
{\em Hodge-K\"ahler manifold} a K\"ahler manifold with K\"ahler form of even
integer cohomology.

\subsection{Special K\"ahler manifolds}
Two questions now arise. The first one is: which
manifolds can be used as scalar manifold for vector multiplets coupled to $N=2$
supergravity? These manifolds are called {\em special (K\"ahler)
manifolds}. The second is: {\em how} can one construct a full supergravity action
starting from a specific special K\"ahler manifold.

In this paper we will be concerned mainly with the first question.
We will propose three definitions of special manifolds and prove
their equivalence. Depending on the context in which supergravity appears one of
them will be more practical than the others.

The second question is briefly touched upon: we show how the spin-1 part of an action
can be constructed.

The previous subsection leads us immediately to the
first definition of a special K\"ahler manifold.

\subsubsection{Definition 1}   \label{ss:def1}

A special K\"ahler manifold\footnote{As in the rigid case, we will always assume a
positive definite \Ka\ metric.\label{fn:posdefmet}} is an $n$-dimensional Hodge-\Ka\
manifold\footnote{i.e. with \Ka\ form of even integer cohomology, see
subsection~\ref{sugract}\label{fn:HodgeKahler}.}
with the following 3~properties.
\begin{enumerate}
\item On every chart there exist complex
projective coordinate functions $Z^I(z)$, where $I=0,\ldots, n$
and a holomorphic function $F(Z^I)$ that is
homogeneous of second degree, such that the \Ka\ potential is
\begin{equation} K(z,\bar z)=-\log\left[i
\bar Z^I \frac{\partial}{\partial Z^I} F(Z) -i
Z^I\frac{\partial}{\partial\bar Z^I}\bar F(\bar Z)
 \right] \ ; \end{equation}
\item
On overlaps of charts $i$ and $j$, the corresponding
functions in property~1 are connected by transition functions
of the following form:
\begin{equation}
\left( \begin{array}{c}
 Z \\ \partial F \end{array}\right)_{(i)} = e^{f_{ij}(z)} M_{ij}
\left( \begin{array}{c}
 Z \\ \partial F\end{array}\right)_{(j)}\ ,
\label{transitionf}
\end{equation}
with  $ f_{ij}$ holomorphic and  $M_{ij} \in \symp{2n+2}$;
\item The transition functions satisfy the cocycle condition on overlap
regions of three charts.
\end{enumerate}

Comparing this definition with the corresponding one in the rigid case
(see section~\ref{rsg:1}), there are several differences.
The $n+1$ coordinates $Z^I$ are projective here (vs. $n$ ordinary
coordinates there), and the expression for the \Ka\ potential is
different. Another difference is that local special geometry involves
{\em local holomorphic} transition functions in the multiplication factor,
vs. constant ones for the rigid case. This is related to the presence of the gauge
field of the local $U(1)$ in the superconformal approach, as should be clear
from section~\ref{sugract}.   \vspace{3mm}

This definition of a special \Ka\ manifold
clearly originates from the construction of an action starting from
a prepotential. Now we would like to get rid of the prepotential in the definition.
The main reason is that it is not invariant under symplectic (duality)
transformations. Actions have been constructed \cite{f0art} for which there does
not even exist a prepotential\footnote{ This case includes
some physically important theories.}.
The following example was given in \cite{f0art}. Take
$n=1$ and start from the prepotential $F(Z)=-i\, Z^0 Z^1$.
Then choose a coordinate $z$ such that $Z^0=1$ and $Z^1=z$.
The symplectic vector in \eqn{transitionf} is then
\begin{equation}
v=\pmatrix{1\cr z\cr -iz\cr -i}\ ,\label{examplev}
\end{equation}
leading to
\begin{equation}
e^{-K}= 2(z+\bar z)\ ;\qquad\partial_z\partial_{{\bar z}}K=
(z+\bar z)^{-2}\ ,     \label{KaSU11U1}
\end{equation}
which is the $SU(1,1)/U(1)$ manifold with positivity domain $\Re
z>0$.
Now perform the symplectic mapping
\begin{equation}
v\rightarrow\widetilde v={\cal S}v
= \pmatrix{1&0&0&0\cr 0&0&0&-1\cr 0&0&1&0\cr 0&1&0&0\cr }
v= \pmatrix{1\cr i\cr -iz\cr z}\ . \label{symplvnoF}
\end{equation}
After this mapping, the transformed vector $\tilde v$ can no longer be
written in the standard form with the function $F$, since
the last two components clearly cannot be written as functions of the first two:
no prepotential $\widetilde F(\widetilde Z^0,\widetilde Z^1)$ exists.
Therefore, after the duality rotation, the model can not be  formulated
directly in  a superspace formulation with a function $F$. We still
have the same \Ka\ manifold \eqn{KaSU11U1}, but with different
couplings of the scalars to the vectors.

In this case it is clear that,
as stressed at the beginning of this subsection, there is no
contradiction with definition~1: our construction made it explicit
that on the
same scalar manifold there is a supergravity action which {\em can} be
obtained from a prepotential: it will be proved below that this is
true in general. Our example
only indicates that the first definition is not
very handy in this case.

The second definition is, as in the rigid case, an intrinsic
symplectic formulation.

\subsubsection{Definition 2}\label{ss:def2}

A special K\"ahler manifold is an $n$-dimensional
Hodge-\Ka\ manifold\footnote{cf. footnotes~\ref{fn:posdefmet}
and~\ref{fn:HodgeKahler} in definition~1.} ${\cal M}$, with the following
2~properties.
\begin{enumerate}
\item There exists a holomorphic $\symp{2(n+1)}$ vector bundle ${\cal H}$ over ${\cal M}$, and a
holomorphic section $v(z)$ of ${\cal L}\otimes{\cal H}$, such that the
\Ka\ form \eqn{defcK} is
\begin{equation}  \label{cond1local}
{\cal K} = -\ft{i}{2\pi}\partial\bar{\partial}\log\left[i\langle {\bar v}
,v\rangle \right]\qquad\mbox{or} \qquad
K=-\log\left[i \langle\bar v,v\rangle\right]\ ;
\end{equation}
Here ${\cal L}$ denotes the holomorphic line bundle over ${\cal M}$
with transition functions as in \eqn{Kahlertransf}, of which the first
Chern class equals the cohomology class of the \Ka\ form.
\item The section $v(z)$ of property~1 satisfies
\begin{equation}
\sinprod{v}{\partial_{\alpha} v} = 0\ .
\label{cond2local}
\end{equation}
\end{enumerate}

This definition is essentially a rewriting of Strominger's \cite{special}
definition, where, however, the second condition, \eqn{cond2local}, is absent.
Comparing with definition~2 of the rigid case in section~\ref{rsg:2}, we notice
that it differs also in that second condition. We now discuss these
differences.

First, let us note that \eqn{cond2local} is a proper equation in the
$\cal L$ bundle, since the \Ka\ covariant derivative of $v$ is
${\cal D}_\alpha v \equiv \partial_\alpha v + (\partial_\alpha K) v$
and the symplectic inner product is antisymmetric. Second, the
condition analogous to \eqn{cond3def2},
\begin{equation}
\sinprod{{\cal D}_\alpha v}{{\cal D}_{\beta} v} = 0\ . \label{cond3local}
\end{equation}
follows from \eqn{cond2local} by taking a (covariant) derivative and
antisymmetrizing. Interestingly, using the invertibility of the \Ka\
metric, the converse is almost true as well: \eqn{cond3local}
{\em implies} \eqn{cond2local} except for $n=1$. This is shown\cite{prtrquat}
by taking a (\Ka\ covariant) antiholomorphic derivative, and noticing
that the curvature is proportional to the invertible \Ka\ metric.
Finally let us remark that in the rigid case it is certainly inappropriate
to impose the analogue of \eqn{cond2local}: it would limit the \Ka\
potential to be homogeneous of second order, which in that context is
an unnecessary restriction.

In the context of Calabi-Yau moduli spaces,
in fact
\eqn{cond3local} is satisfied automatically (see section~\ref{ss:CY}). To show
the necessity of this condition in the general context we have constructed
counterexamples, given in detail
in  appendix~\ref{counterexample}. They show that without
these conditions one could have manifolds which are not special
according to definition~1, and to the known $N=2$
supergravity constructions.
We come back to the interpretation of the constraints at the end of
this section.

\paragraph{Remark 1.} \label{remark1}

The (local) constraints \eqn{cond1local},
\eqn{cond2local} (and \eqn{cond3local}) can be equivalently
formulated in terms of a different bundle, with section
$V\equiv e^{K/2}v$ where  $K$ is the \Ka\ potential.
This $V$ is the one appearing in \eqn{constraint}.
With
\begin{eqnarray}
U_{\alpha}\equiv
\Dee{\alpha}V \equiv \dee{\alpha}V+\ft{1}{2}(\dee{\alpha}K)V&
;&\qquad \Dee{\bar\alpha}V \equiv \dee{\bar\alpha}V
-\ft{1}{2}(\dee{\bar\alpha}K)V  \nonumber\\
\bar U_{\bar \alpha}\equiv
\Dee{\bar \alpha}\bar V \equiv \dee{\bar \alpha}\bar V+
\ft{1}{2}(\dee{\bar \alpha}K)\bar V&
;&\qquad \Dee{\alpha}\bar V \equiv \dee{\alpha}\bar V
-\ft{1}{2}(\dee{\alpha}K)\bar V
\ ,    \label{defcovderV}
\end{eqnarray}
the constraints are:
\begin{enumerate}
 \item $\langle V,\bar V\rangle =i     \ ; $  \label{eisVVbar}
 \item $\Dee{\bar \alpha}V=0\ ;  $ \label{eisDabarV}
 \item $\langle V,U_{\alpha}\rangle =0 \ ;$ \label{eisVUa}
 \item $\langle U_{\alpha},U_{\beta}\rangle=0\ . $ \label{eisUaUb}
\end{enumerate}
As explained above, given \ref{eisVVbar} and \ref{eisDabarV},
constraint~\ref{eisVUa} implies \ref{eisUaUb},
and \ref{eisUaUb} implies \ref{eisVUa} unless $n=1$ (in which case
constraint~\ref{eisUaUb} is empty).

We now turn to the proof of the equivalence of our two definitions.
In this section we just give the outline in order to make clear
the structure of the proof. The lemmas that constitute the different steps
are proved in detail in  appendix~\ref{1=2lemmas}.

We will formulate the proof mostly in terms of the section $V$, as in
remark~1.
Before embarking, let us make explicit the
inner products of the symplectic vectors $V$, $U_\alpha$ and their
complex conjugates. Taking  a covariant
derivative of condition~\ref{eisVVbar} one immediately gets
\begin{equation}
\langle U_{\alpha},\bar V\rangle=0\ ;\qquad
\langle\bar  U_{\bar \alpha},V\rangle=0\ .  \label{UbarV0}
\end{equation}
Noticing that the curvature in the bundle is essentially the \Ka\ form:
\begin{equation}
\left[{\cal D}_\alpha,{\cal D}_{\bar \beta}\right]V=
-g_{\alpha\bar \beta} V \ ,
\end{equation}
we can take a further covariant derivative of \eqn{UbarV0} to obtain
\begin{equation}
 \langle U_{\alpha},\bar  U_{\bar \beta}\rangle
 =-ig_{\alpha\bar \beta} \ . \label{UbarUg}
\end{equation}

\paragraph{Outline of the proof.}

The conditions of definition~1 clearly imply those
of definition~2 (see \cite{prtrquat}). So we only have to
prove the converse.

The (global) section $v$ in definition~2 is, in each patch, represented
by a vector, and the transition between patches is made with
transition functions that clearly satisfy the
last two conditions of definition~1. For definition~1 it is in addition
necessary that these vectors are of the form $\left( \begin{array}{c}
Z \\ \partial F \end{array}\right)$ for some function $F$.
We will show how to construct, starting from an arbitrary vector $v$
that satisfies condition~\ref{cond2local}, a vector of that form,
using constant \Sp{2n+2}-transformations. The transition functions
then keep the form required in \eqn{transitionf}.

It is thus {\em sufficient to prove} that given a section  $V=\left(
\begin{array}{c} X^I\\F_I\end{array}\right)$  satisfying
the conditions of remark~1, there exists an
\Sp{2n+2}-transformation which transforms $V$ into a
vector  $\tilde V=\left(
\begin{array}{c}\tilde X^I\\ \tilde F_I\end{array}\right)$
with the property that the transformed $\tilde F_I$ is the derivative of
a second degree homogeneous holomorphic
function:
$\tilde F_I= \frac{\partial}{\partial\tilde X^I}\tilde F(\tilde X)$.

Since the metric $g$ is non-degenerate,
lemma~\ref{lemma551} implies that
\[
 {\rm rank}\mtrx{f_\alpha^I}{h_{\alpha I}}{X^I}{F_I}=n+1\ ,
\]
where we have introduced notations for the components of $U_\alpha$:
\begin{equation}
U_\alpha=\pmatrix{f_\alpha^I\equiv \Dee{\alpha}X^I\cr h_{\alpha I}\equiv
\Dee{\alpha}F_I}\ .
\end{equation}
According to lemma~\ref{lemma553} there exists an  $S\in\Sp{2n+2}$ such that
\[SV\equiv \tilde V= \left(\begin{array}{c}\tilde X^I\\ \tilde F_I
\end{array}\right)\ ,\]
 with
 \begin{equation}
 \det \left(\begin{array}{c}\tilde f_\alpha^I\\
\tilde X^I\end{array}\right)\ne 0\ .\label{invXF}
\end{equation}
Using lemmas \ref{lemma555} and \ref{lemma556} we find a
function $\tilde F(\tilde X)$
such that $\tilde F_I=\frac{\partial\tilde F}{\partial\tilde X^I}$.
Note that we have made use of the constraints \eqn{eisUaUb}, \eqn{eisVUa}
and \eqn{eisDabarV} of remark~1, which are \Sp{2n+2} invariant
and are thus satisfied by $\tilde V$ too. This completes the
proof of the equivalence of both definitions.

The equivalence of the two definitions of a special \Ka\ manifold
does {\em not} imply that the formulation of the corresponding
supersymmetric theories without a prepotential may be discarded.
It only
means that we can always describe the geometry of
the scalars by means of a prepotential. It is not
(necessarily) so that the symplectically rotated supergravity
theory is equivalent to the original one. Especially for
gauged (non-abelian) theories some interesting phenomena
have been discovered:
\begin{itemize}
\item It is possible that symplectically related abelian
theories allow different gauge groups.\cite{f0art}
\item In theories which have a prepotential it is
impossible to partially break $N=2$ supersymmetry to $N=1$
by means of Fayet-Iliopoulos terms; this phenomenon does occur in
some theories without a prepotential.\cite{N=2N=1}
\end{itemize}
\paragraph{The kinetic term of the vectors.}

Having investigated the scalar sector, we now come back
to the question of how to construct the remaining part
of the action of supergravity coupled to vector multiplets.
The complete action constructed in \cite{dWLVP} was rewritten in
terms of these symplectic building blocks in \cite{f0art,ItalianN2}.
The tensor ${\cal N}$ has a direct physical interpretation as its
expectation value gives the coupling constants in the kinetic term of
the vectors, see \eqn{genL01}. Its symmetry was also the essential
ingredient in restricting transformations between Bianchi identities
and field equations to symplectic ones. Similarly, the extra constraints
necessary for special \Ka\ manifolds can be seen to originate from
this requirement of symmetry. Of course,
if a prepotential exists for $v$ we can construct ${\cal N}$ via formula
\eqn{NvanF}. This matrix is explicitly
symmetric. Moreover it has been mentioned already that the negativeness of
its imaginary part follows from the positivity of kinetic terms of
scalars and vectors, as proven in \cite{BEC}.
These properties are preserved under symplectic transformations
(corollary~\ref{gevolg}) and as we already know
that in any case we can find a symplectic transformation to
a formulation where the prepotential exist, we could stop here.
However, it is useful to have a definition which starts from the
symplectic vector $v(z^{\alpha})$, and see how constraints of
special geometry can be understood as requirements on ${\cal N}$.

In order to define \N\ in the general case we need the invertibility of
\begin{equation}
\left( \begin{array}{cc} f_\alpha^I&\bar X^I
\end{array}\right)\ .  \label{invbarVU}
\end{equation}
This is proven in corollary~\ref{cor:gevolg2}, whose conditions
correspond again to the positivity of the spin 2 and spin 0 kinetic
terms. Note that the lemma~\ref{lem:invbarVU} which led to that
corollary was derived under a weaker condition than
condition~\ref{eisVUa}. We will come back to this remark below.

Then we can define as in equation \eqn{NXY}:
\begin{equation}  \label{defvanN}
{\cal N}_{IJ}\equiv \left( \begin{array}{cc} \bar h_{\bar\alpha I}&F_I
\end{array}\right)  \left( \begin{array}{cc} \bar
f_{\bar\alpha}^J&X^J
\end{array}\right)^{-1}   \ ,  \label{sympldefcN}
\end{equation}
which, as mentioned below \eqn{NXY}, transforms under symplectic
transformations as in \eqn{tildecalN}.
If a prepotential exists this definition coincides with \eqn{NvanF}
\cite{prtrquat}.

The equation
\begin{equation}
i\pmatrix{\langle U_\alpha ,\bar U_{\bar \beta}\rangle &
\langle U_\alpha ,V\rangle
\cr \langle \bar V,\bar U_{\bar \beta}\rangle&\langle \bar V,V\rangle }=
\pmatrix{f_\alpha^I\cr \bar X^I}i({\cal N}- {\cal N}^{\dagger})_{IJ}
\pmatrix{\bar f_{\bar \beta}^J& X^J}        \label{ImcN}
\end{equation}
follows from the definition of ${\cal N}$, and thus implies the
negative definiteness of the antihermitian part of ${\cal N}$ as a
consequence of the positivity of the matrix on the left hand side.
We can here make the same remark about the necessity or weakening of
the condition~\ref{eisVUa} as after \eqn{invbarVU}.

The symmetry of ${\cal N}$ follows from a similar equation:
\begin{equation}
\pmatrix{ f_{ \alpha}^I\cr  \bar X^I}(\bar {\cal N}-\bar {\cal N}^T)_{IJ}
\pmatrix{ f_{ \beta}^J&\bar  X^J} =
\pmatrix{\langle U_{ \alpha} , U_{ \beta}\rangle &
\langle U_{ \alpha} ,\bar V\rangle
\cr \langle \bar V, U_{ \beta}\rangle&0 }\ . \label{symcNsympl}
\end{equation}
The off-diagonal elements of the right hand side are zero due to
\eqn{UbarV0}, a consequence of condition~\ref{eisVVbar}. The upper
left entry is zero due to condition~\ref{eisUaUb}. The invertibility
of \eqn{invbarVU} implies the symmetry of ${\cal N}$.

\paragraph{Interpretation of the constraints.}

Now we are in a position to interpret our extra constraints in the
second definition of special geometry. The condition~\ref{eisDabarV}
is essential for the holomorphic structure. The condition~\ref{eisVVbar}
is a normalization which is possible if the spin-2 kinetic terms are
positive definite. It corresponds to the conventional parametrization
of the spin-2 action in the `Einstein frame'.
Similarly the form of \eqn{defcovderV}, or equivalently \eqn{UbarUg},
corresponds to the scalar kinetic terms. For the spin-1 kinetic
terms (i.e. first to be able to define ${\cal N}$ uniquely, and also
explicitly in the expression for $\Im{\cal N}$ which should be
negative definite for positive definite kinetic terms), we need the
positivity of the matrix \eqn{ImcN}. Finally, the
condition~\ref{eisUaUb} is necessary for the symmetry of ${\cal N}$,
as seen explicitly in \eqn{symcNsympl}.

The above arguments still did not lead to condition~\ref{eisVUa}, but
to the weaker condition of the positivity of \eqn{ImcN}. It is clear
that if condition~\ref{eisVUa} is satisfied, that matrix is positive.
The relaxation of the constraint is only relevant for $n=1$. Indeed
\cite{prtrquat}, from \eqn{defcovderV} ($U$ has the same \Ka\ weight
as $V$) one derives
\begin{equation}
{\cal D}_{\bar \beta}U_\alpha=g_{\alpha\bar \beta}V\ .
\end{equation}
Then applying $g^{\alpha\bar \gamma}{\cal D}_{\bar \gamma}$ to
condition~\ref{eisUaUb} gives condition~\ref{twee} if $n>1$.

Remains $n=1$. In appendix~\ref{counterexample} we give two examples
which prove that the extra constraint is relevant for $n=1$, i.e.
that relaxing this constraint effectively enlarges the allowed
symplectic sections. In a first example, it is shown that having only
\eqn{defcovderV} (or \eqn{UbarUg}) and conditions~\ref{eisVVbar}
and~\ref{eisDabarV}, would allow a larger class of \Ka\ manifolds,
i.e. \Ka\ manifolds which are not allowed with
condition~\ref{eisVUa}. In that example the matrix \eqn{ImcN} is not
postive definite.
A second example (related to the simple
example we gave at the end of section~\ref{ss:def1}) shows a
symplectic section for which \eqn{ImcN} is positive definite, but the
condition~\ref{eisVUa} is violated. The \Ka\ manifold is the same as
the one of \eqn{KaSU11U1}, but the symplectic section is different
and is {\em not} related to that in \eqn{examplev} by a symplectic
rotation, as opposed to the one of \eqn{symplvnoF}. We discuss in
appendix~\ref{app:example2} in how far this model violates the
equations occuring in the constructions of $N=2$ supergravity models.

In view of the fact that the constraint \eqn{cond3local} has a direct
interpretation in terms of \N, whereas \eqn{cond2local} does not, one
may wonder if it would be more appropriate to figure the latter
together with the positivity of \eqn{ImcN} in
definition~2 of special \Ka\ manifolds. This would increase also the
similarity with the rigid case. However, there exist at present no
actions of $N=2$ supergravity coupled with vector multiplets which
give rise to symplectic sections which do not satisfy our
constraints. So, the relaxation would not be in agreement with our
general strategy that the sections satisfying the definitions should
allow such an $N=2$ supergravity action. Also, we would not be able
to show the
equivalence with the first definition, which was directly related to a
supergravity action.

\paragraph{Invertibility of $(n+1)\times(n+1)$ matrices and existence
of the prepotential.}

We draw attention to the  different status of the invertibility
on the matrices in  \eqn{invXF} and \eqn{invbarVU}.
The latter is always invertible, while the invertibility of the former
is a criterion
for the existence of a prepotential in this symplectic basis. To
illustrate this, consider the example \eqn{symplvnoF}. Taking the
covariant derivatives, the matrix  \eqn{invbarVU}  is
\begin{equation}
\left( \begin{array}{cc} f_\alpha^I&\bar X^I \end{array}\right)=
e^{K/2}\pmatrix{-\frac{1}{z+\bar z}&1\cr -\frac{i}{z+\bar z}&-i}\ ,
\end{equation}
which is invertible. If the second column had been $X$ rather than $\bar
X$, the sign change in the last entry would have made this matrix
non-invertible, reflecting the non-existence of the prepotential for
this symplectic section.
\subsubsection{Definition 3}
There is a third definition (inspired by
\cite{CDFLL,f0art,prtrquat}),
which is analogous to the third definition in the rigid case.
Take a complex manifold \M. Suppose we have in every chart a $2(n+1)$ component
vector $V(z^{\alpha},\bar z^{\alpha})$ such that on overlap regions there are
transition functions
 of the form \[e^{\frac{1}{2}(f(z^{\alpha})
-\bar f(\bar z^{\alpha}))}\,S\ ,\] with $f$ a holomorphic function and
$S$ a constant \Sp{2(n+1)}  matrix. (These transition functions
have to satisfy the cocycle condition.) Take a $U(1)$ connection of the form
$\kappa_{\alpha}\,dz^{\alpha}+
\kappa_{\bar \alpha}\,d\bar z^{\alpha}$ with
\begin{equation}
\kappa_{\bar\alpha}= -\overline{\kappa_{\alpha}}\ ,  \label{defkappabar}
\end{equation}
under which $\bar V$ has opposite
weight as $V$.
Denote the covariant derivative by ${\cal D}$:
\begin{equation}
\begin{array}{ll}
U_\alpha\equiv {\cal D}_{\alpha} V\equiv
\partial_{\alpha} V+\kappa_{\alpha} V
\ ;\qquad&
{\cal D}_{\bar \alpha}V\equiv
\partial_{\bar\alpha} V+\kappa_{\bar\alpha} V\\
\bar U_{\bar \alpha}\equiv {\cal D}_{\bar\alpha}\bar V\equiv
\partial_{\bar\alpha}\bar V-\kappa_{\bar\alpha}\bar V
\ ;\qquad&
{\cal D}_{\alpha}\bar V\equiv
\partial_{\alpha}\bar V-\kappa_{\alpha}\bar V \ .
\end{array}            \label{covder3}
\end{equation}
We impose the following conditions:
\begin{enumerate}
 \item $ \langle V,\bar V\rangle =i     \ ;$                 \label{een}
 \item $ \Dee{\bar \alpha}V=0           \ ;$                 \label{vier}
 \item $ \Dee{[\alpha}U_{\beta]}=0      \ ;$                 \label{zes}
 \item $ \langle V,U_{\alpha}\rangle =0 \ .$                 \label{twee}
\end{enumerate}
Define
\begin{equation}  \label{defg}
 g_{\alpha\bar\beta}\equiv i\langle U_{\alpha},\bar U_{\bar\beta}\rangle \ ,
\end{equation}
where $\bar U_{\bar\beta}$ denotes the complex conjugate of $U_{\alpha}$.
If this is a positive-definite metric,
\M\ is called a special \Ka\ manifold.

We now prove the equivalence of this definition with definition~2.
\begin{proof}
The second definition provides
all the equations in remark~1, which appear here as
conditions~\ref{een},~\ref{twee}~and~\ref{vier}.
The values of $\kappa_\alpha$ and $\kappa_{\bar \alpha}$ can be
obtained by comparing \eqn{defcovderV} and \eqn{covder3}, and they
satisfy \eqn{defkappabar}. Condition~\ref{zes} is then also easily
checked. This shows that a manifold satisfying definiton~2 also
satisfies definition~3.

We now turn to the converse.
This requires a proof of the existence of the symplectic section mentioned in
definition~2, from the constraints in definition~3.
{}From condition~\ref{zes}  we get
\[0=[\Dee{\alpha},\Dee{\beta}]V=2(\dee{[\alpha}\kappa_{\beta]})V\ ,\]
and taking an inner product with $\bar V$ implies that
$\dee{[\alpha}\kappa_{\beta]}=0$,
so a function $K'$ exists for which
\begin{equation}
 \label{kappa}
 \kappa_{\alpha}=\ft{1}{2}\dee{\alpha}K'                        \ .
\end{equation}
With \eqn{defkappabar}, this implies
\begin{equation}
 \label{kappabar}
 \kappa_{\bar\alpha}=-\overline{\kappa_{\alpha}}=
 -\ft{1}{2}\dee{\bar\alpha}\bar K'\ .
\end{equation}
Applying derivatives on condition~\ref{een} we obtain \eqn{UbarV0}.
Also
\begin{eqnarray}
0&=&\partial_{\bar\beta} \langle\bar V,U_\alpha\rangle
=\langle\bar U_{\bar\beta},U_{\alpha}\rangle+\langle\bar V,
\Dee{\bar\beta}\Dee{\alpha}V\rangle
=ig_{\alpha\bar\beta}+2i\dee{[\alpha}\kappa_{\bar\beta]} \nonumber\\
&&\Rightarrow\  \label{gkaehler}
 g_{\alpha\bar\beta}=- 2\dee{[\alpha}\kappa_{\bar\beta]}
 =\dee{\alpha}\dee{\bar\beta}(\RE K')                          \ ,
\end{eqnarray}
and the real part of $K'$ is thus the \Ka\ potential.
 Condition \ref{vier} and equation \eqn{kappabar} lead us to the
conclusion that the holomorphic $v$ of definition~2 can be obtained
from
\begin{equation}
V=e^{\bar K'/2}v=e^{-i\,\IM K'/2}e^{\RE K'/2}v\ .
\end{equation}
This implies that \M\ is special \Ka\
according to the first two definitions.
\end{proof}

Some further remarks:
\begin{itemize}
\item
the vector
\[V'\equiv e^{i\IM K'/2}V=e^{\RE K'/2}v\]
satisfies the conditions~\ref{een} till~\ref{vier}
of remark~1, page~\pageref{remark1}.
\item The remarks about possible replacement of the
condition~\ref{twee} by
\begin{equation}
\mbox{Condition \ref{twee}' : } \langle
U_{\alpha},U_{\beta}\rangle=0\ ,
\label{drie}
\end{equation}
together with, for $n=1$ the positive definiteness of \eqn{ImcN},
apply here as well.
\item Defining
\begin{equation}
C_{\alpha\beta\gamma}=-i\langle {\cal D}_\alpha U_\beta,
U_\gamma\rangle\ ,
\end{equation}
we can see that it is completely symmetric with the help of
\eqn{drie} and condition~\ref{zes}.
\begin{equation}
{\cal D}_\alpha U_\beta =C_{\alpha\beta\gamma}  \bar U^\gamma\qquad
\mbox{with}\qquad
\bar U^\alpha\equiv g^{\alpha\bar \beta}\bar U_{\bar \beta}\ .
\label{alterndefC}
\end{equation}
\item Rather than imposing condition~\ref{twee}, in the literature
\cite{CDFLL,f0art,ItalianN2} one often imposes \eqn{alterndefC},
together with the symmetry of $C_{\alpha\beta\gamma}$%
\footnote{One only has to require that
$C_{\alpha\beta\gamma}$ be symmetric in the first two indices (the
complete symmetry then follows later), which corresponds to our
condition~\ref{zes}.}.
This alternative is nearly equivalent to the conditions that we have imposed,
but not quite.
{}From the covariant derivatives of
condition~\ref{een}, using condition~\ref{vier}
one derives consecutively $\langle U_\alpha,\bar V\rangle=0$,
$\langle {\cal D}_\beta U_\alpha,\bar V\rangle=0$, and
$\langle {\cal D}_\beta U_\alpha,\bar U_{\bar \gamma}\rangle=0$.
Combining these  with \eqn{alterndefC} leads to
\begin{equation}
C_{(\alpha\beta)\gamma}g^{\gamma\bar \gamma}\langle\bar U_{\bar \gamma},
\bar V\rangle=0  \qquad\mbox{and}\qquad
C_{(\alpha\beta)\gamma}g^{\gamma\bar \gamma}\langle\bar U_{\bar \gamma},
\bar U_{\bar \delta}\rangle =0 \ .
\end{equation}
If the $(n(n+1)/2)\times n$ matrix
$C_{(\alpha\beta)\gamma}$ is of rank $n$,
one can deduce from these equations that the symplectic inner products
 vanish. If this is the case, starting from \eqn{alterndefC}
is equivalent to the  condition~\ref{twee} that we imposed
\footnote{A similar remark could be made for the rigid
case, replacing \eqn{rsgdefuu} with \eqn{alterndefC} if $C$ satisfies
the same non-degeneracy condition.}.
There are, however, manifolds for which this is not the
case: the so-called 'minimal coupling models' \cite{mincoupl},
where $F$ is a quadratic polynomial of the $X$, provide an example
since they have
$C_{(\alpha\beta)\gamma}=0$. For $n=1$, this
is in fact the manifold that we used as an example before.
\end{itemize}
\paragraph{Matrix formulation.}

Just as in the rigid case, the constraints of definition~3
\cite{CdAF,CDFLL,f0art,prtrquat,ItalianN2} can be
summarised in a matrix form, which acquires a geometric significance
in the realisation of these manifolds as moduli spaces of Calabi-Yau
manifolds. We define the $2(n+1)\times 2(n+1)$ matrix
\begin{equation}
\cal V = \pmatrix{\bar V^T\cr  U^T_\alpha \cr V^T\cr
\bar  U^{\alpha\,T} \cr}\ .\label{def3matrix}
\end{equation}
All the inner products and differential equations satisfied by these
quantities can be summarized
in the relations
\begin{eqnarray}
&&{\cal V}\,\Omega \,{\cal V}^{\rm T} = i\Omega\ ;\nonumber\\
&&{\cal D}_\alpha {\cal V}= {\cal A}_\alpha {\cal V}\ ;
\qquad{\cal D}_{\bar \alpha }
{\cal V}={\cal A}_{\bar \alpha }{\cal V}\,. \label{flatness}
\end{eqnarray}
with
\begin{equation}
{\cal A}_\alpha  = \pmatrix{0&0&0&0 \cr \noalign{\vskip1mm}
              0&0&0 &C_{\alpha \beta \gamma }\cr \noalign{\vskip1mm}
              0&\delta_\alpha ^\gamma&0&0\cr \noalign{\vskip1mm}
              \delta^\beta _\alpha& 0&0&0\cr }\ .
\end{equation}
The covariant derivatives contain the \Ka\ connection, and when
acting on $U_\alpha$, also Levi-Civita connection as in
\eqn{LeviCivita}. As for the rigid case one proves (see e.g.
\cite{prtrquat}) that $C$ is a symmetric tensor, covariantly
holomorphic, and using commutators of covariant derivatives, one
obtains the curvature formula
\begin{equation}
R^\alpha {}_{\beta \gamma}{}^\delta
=2\delta^\alpha_{(\beta}\delta^\delta_{\gamma)}-
C_{\beta \gamma \epsilon}\bar C^{\alpha\delta\epsilon}\label{Rlocal}
\ .
\end{equation}
It is clear that the conditions \eqn{flatness} are an over-complete
set of requirements, and therefore a definition based on them does
not seem to be economical.

\subsection{Calabi-Yau moduli spaces}\label{ss:CY}
It has been discovered \cite{Seiberg,CecFerGir,FerStroCand,special}
that moduli spaces of Calabi-Yau \ threefolds
present themselves as natural candidates for special \Ka\ manifolds.
As will be argued below, the third
formulation of special manifolds is tailored for making the
connection.

A Calabi-Yau threefold  is a compact \Ka\ manifold of complex dimension 3 whose
first Chern class vanishes.
The Hodge diamond of a Calabi-Yau threefold exhibits a lot of symmetry. For
example, if the Euler character is non-zero, the diamond has the following
form:
\begin{center}
\begin{tabular}{ccccccc}
 & &  & $h^{00}=1$      &   &  & \\
 & &0 &        & 0 &  & \\
 &0&  &$h^{11}=m$&   & 0& \\
$h^{30}=1$& &$h^{21}=n$&  &$h^{12}=n$&&$h^{03}=1$\\
 &0&  &$h^{22}=m$&   & 0& \\
 & &0 &        & 0 &  & \\
 & &  & $h^{33}=1$      &   &  &
\end{tabular}
\end{center}
The matrix \eqn{def3matrix} will be related to the period matrix
of integrals of $(3,0), \, (2,1), \, (1,2)$ and $(0,3)$ forms (the
rows of the matrix) over 3-cycles. There are $2(n+1)$ homologically
different 3-cycles in the CY manifold. One chooses a
canonical basis with intersection numbers, as for the 1-cycles on
Riemann surfaces, \eqn{intersections}. Again, symplectic rotations correspond
to changes of this canonical homology basis, and the restriction to
integers is natural.

E.g. take the unique (up to normalization) holomorphic threeform: $\Omega\in
H^{(3,0)}$. We choose $\Omega$ to depend holomorphically on the
moduli. We identify
\[
v \equiv { \int_{A^I}\Omega \choose \int_{B_I}\Omega}
\ .
\]
The vector $v$ depends on the moduli
of the Calabi-Yau manifold. At this
point we could still multiply $\Omega$ by a holomorphic function of
the moduli. Such a transformation will correspond to the \Ka\
transformations in special geometry.

On the Calabi-Yau manifold there is also the integral formula,
similar to \eqn{integralformula}, now for 3-forms and 3-cycles.
This allows us to identify the symplectic inner product of
symplectic vectors, formed by periods of 3-forms, with the integral of
the exterior product of the forms over the full Calabi-Yau space.
With a suitable orientation $-i\,d^3x\,d^3\bar x>0$ (where $x$ are
the holomorphic coordinates on the Calabi-Yau surface),
and we have
\begin{equation}
-i\langle v,\bar v\rangle = -i\int_{CY}\Omega\wedge \bar \Omega >0\ ,
\end{equation}
and we can thus identify the latter as $\exp (-K)$.

Small variations of a $(p,q)$ form can give at most $(p\pm 1,q\mp 1)$
forms. Applied to the unique (3,0) form this gives \cite{FerStroCand}
\begin{equation}
\partial_\alpha \Omega= \Omega_\alpha -k_\alpha \Omega  \ ,
\end{equation}
where $\Omega_\alpha$ are (2,1) forms and $k_\alpha$ functions of the
moduli. The integral formula implies that only inner products of
periods of $(p,3-p)$ with $(3-p,p)$ forms are non-vanishing.
Therefore one can already see that \eqn{cond2local} and \eqn{cond3local}
are satisfied trivially.
(This observation can explain their absence in \cite{special}.)
One then also obtains
\begin{equation}
k_\alpha =-\frac{\langle \partial_\alpha \Omega,\bar \Omega\rangle }
{\langle  \Omega,\bar \Omega\rangle}=\partial_\alpha K\ .
\end{equation}
This allows us to identify (integrals over canonical basis of 3-cycles)
\begin{equation}
V=e^{K/2}\int \Omega\ ;\qquad U_\alpha=\left( \partial_\alpha +
\ft12(\partial_\alpha K )\right) V=e^{K/2}\int \Omega_\alpha\ .
\end{equation}
If we then take $i\langle U_\alpha,\bar U_{\bar \beta}\rangle $, we
still have the order of the $dx$ and $d\bar x$ in the integral that
guarantees this to be a positive definite matrix, and thus all
requirements for special geometry are fulfilled.

We thus obtain the following identifications:
\begin{equation}
V=e^{K/2}\int \Omega^{(3,0)}\ ;\  U_\alpha=e^{K/2} \int \Omega_\alpha^{(2,1)}
\ ;  \
\bar U_{\bar \alpha}= e^{K/2}\int \Omega_{\bar \alpha}^{(1,2)}
\ ; \
\bar V=e^{K/2}\int \Omega^{(0,3)}\ .
\end{equation}
The relations satisfied by the derivatives of these periods, the
so-called Picard-Fuchs equations, take the form of the
defining equations of special geometry.

It is now interesting to translate our results on the invertibility
of $(n+1)\times(n+1)$ matrices to these periods. First of all, the
conditions of lemma~\ref{lem:invbarVU} are satisfied: the off-diagonal
elements are zero because after using the integral formula one
obtains products of $(3,0)$ and $(2,1)$. The diagonal elements are
positive because of the order of $dx$ and $d\bar x$ factors as
mentioned above. Therefore the lemma leads now directly to the
invertibility of the $(n+1)\times(n+1)$ matrix formed by integrating
the $(3,0)$ and $(1,2)$ forms over the $A$-cycles, {\em for every
choice of non-intersecting independent $A$-cycles}. Furthermore
lemma~\ref{lemma556} implies that if we have chosen $A$-cycles such
that the integrals of the $(3,0)$ and $(2,1)$ forms over the
$A$-cycles gives an invertible matrix, then a prepotential exists.
Lemma~\ref{lemma553} says that one can always choose such a basis of
cycles. However, the statement often made for Calabi-Yau manifolds
\cite{FerStroCand} that with
\begin{equation}
Z^I\equiv \int_{A^I}\Omega^{(3,0)} \ ;\qquad
{\cal G}_I\equiv \int_{B_I}\Omega^{(3,0)}\ ,  \label{zcG}
\end{equation}
the complex structure of the Calabi-Yau manifold is entirely
determined by the $z^I$ (as projective coordinates), so that ${\cal
G}_I= {\cal G}_I(z)$ can thus be violated for a particular choice of
the cycles. That is the content of the possibility of absence of
prepotential, found in \cite{f0art}, translated to the Calabi-Yau
context. The choice of cycles can get a physical meaning in the
applications in string theory.

\section{Remark and conclusions}\label{ss:concl}

We have given several equivalent definitions of special geometry.
For the rigid as well as for the local case, one can define special
geometry from a prepotential. However, a more useful definition
starts from a symplectic bundle.
The symplectic transformations, inherited from the
dualities on vectors, are a crucial aspect for the scalar manifolds.
A few extra constraints have
to be imposed in order that this symplectic bundle can lead to a special
manifold. It is in these extra constraints that earlier proposals
 for a 'coordinate free' definition were incomplete.
The missing equations are mostly related to the condition of
symmetry~\cite{prtrquat} of the matrix ${\cal N}_{IJ}$,
which is a key ingredient in the discussion of symplectic transformations.

It is always immediate to go from the formulation with a prepotential
 to the symplectic one. In rigid special geometry, the existence of a
prepotential can be seen for any symplectic bundle which satisfies
the requirements in our definition. In the local case, this
connection can be less straightforward. If the metric is positive
definite it can in general be shown that the $(n+1)\times(n+1)$ matrix
\begin{equation}
\left( \begin{array}{cc} \Dee{\alpha} X^I&\bar X^I
\end{array}\right)  \label{invbarVUc}
\end{equation}
is invertible. However, a prepotential only exists for the symplectic
section if
\begin{equation}
\left( \begin{array}{cc} \Dee{\alpha} X^I& X^I
\end{array}\right)  \label{invVUc}
\end{equation}
is invertible\footnote{In rigid supersymmetry these are $n\times n$
matrices, the last column is not there and the index $I$ is replaced
by $A=1,\ldots ,n$. Thus the two matrices are then identical.}.
The interesting supergravity models without
prepotential are thus those for which the latter matrix has a zero
mode. However, we have shown in general that one can then always make
a symplectic transformation to a basis in which this matrix becomes
invertible, and thus a prepotential does exist. As the scalar
manifold is invariant under such a symplectic transformation, we can
say that any special manifold can be obtained from a prepotential.

Special geometry is sometimes
defined \cite{CDFLL,fresoriabook,ItalianN2}
by giving the curvature formulas \eqn{Rrigid} for the rigid
case or \eqn{Rlocal} for the local case,
for a symmetric tensor $C$, with
${\cal D}_{\bar \alpha}C_{\alpha\beta\gamma} =0$ and
${\cal D}_{[\alpha}C_{\beta]\gamma\delta}=0$.
Whereas these equations are always valid, we have not investigated
(and know of no proof elsewhere)
in how far they constitute a sufficient condition.

We have also shown how certain moduli spaces of Riemann surfaces and
moduli spaces of Calabi-Yau manifolds satisfy the definitions
we have given. It is the reappearance of special geometry in these contexts
which has led to recent new applications.
As Riemann and Calabi-Yau surface theory and $N=2$ field theories
are related through the concept of special geometry, questions on
one side can find an answer on the other side.
We have seen that the connection of (local) special geometry to the
moduli spaces of Calabi-Yau manifolds is rather straightforward.
These spaces can always be endowed with a special \Ka\ metric.
However, for moduli spaces of Riemann surfaces  the constraint may
necessitate a proper identification of some particular subspace in
the full moduli space. We have shown in section~\ref{ss:RS} how our
definition can be satisfied, but a better understanding of the
geometrical significance of these subsets of moduli spaces would be
welcome.

For Calabi-Yau moduli spaces, the statements about the matrices
\eqn{invbarVUc} and \eqn{invVUc} imply, that for any choice of basis
of non-intersecting (`$A$') cycles, the matrix of periods of the
$(3,0)$ and $(1,2)$ forms over these cycles is invertible. On the
other hand, there is always a basis of cycles such that the integrals
of the $(3,0)$ and $(2,1)$ forms over the $A$ cycles is invertible.
A prepotential, or even a dependence ${\cal G}_A(z)$ in the
sense of \eqn{zcG}, exists if and only if such a basis is chosen.

\medskip
\section*{Acknowledgments.}

\noindent
We thank P. Candelas, B. de Wit, P. Fr\`e and P. West for discussions.
This work was supported by
the European Commission TMR programme ERBFMRX-CT96-0045. Part of this
work was performed during a visit of A.V.P. at the Newton Institute in
Cambridge.
\newpage

\appendix
\section{General theorems on symplectic transformations}
\label{app:proofssymp}
First  we will prove here some general properties of symplectic
transformations. One important lemma (lemma~\ref{lemma553})
says that for a rank $m$ matrix \eqn{defmatV} there
 always exist symplectic
transformations $\tilde V={\cal S}V$ such that the upper half of $\tilde V$
becomes invertible.
The proof leads immediately to property~\ref{prop:sympfact}, which
says that
all symplectic transformations can be written as a product of a finite
number of symplectic matrices of the following form:
\begin{equation}
L=\pmatrix{ A&0\cr C&{A^{-1T}} }\ ;\qquad
S=\pmatrix{0&\unity\cr -\unity&0} \ .
\end{equation}

If $X$ is invertible we can define ${\cal N}$ as in \eqn{NXY},
transforming as in \eqn{tildecalN}. The invertibility of $X$, the
symmetry of ${\cal N}$ and the negative definiteness of its imaginary
part are preserved by these symplectic transformations
(theorem~\ref{pq}). It is also to be remarked that
any symmetric matrix with negative definite imaginary part is
invertible (lemma~\ref{N}). Any such matrix ${\cal N}$ can of course
be written in a form \eqn{NXY}, taking $X=\unity $, and for such a
matrix $(A+B{\cal N})$ is invertible and the matrix $\tilde{\cal N}$
in \eqn{tildecalN} is thus symmetric with negative definite imaginary
part (corollary~\ref{gevolg}). In fact,
\begin{equation}
\Im \tilde{\cal N}=(A+B\bar {\cal N})^{-1\, T}(\Im {\cal N})(A+B{\cal
N})^{-1}
\end{equation}
 already shows the preservation of the signature.

It is useful to realise first that for a general
symplectic matrix
\begin{equation}
{\cal S}=\pmatrix{A&B\cr C&D\cr}  \ ,
\end{equation}
the defining relations \eqn{symplABCD} imply that
the following matrices are all symmetric  $A^T C$, $B^T D$, $CA^{-1}$,
$BD^{-1}$, $A^{-1}B$, $D^{-1}C$, $AB^T$, $DC^T$ (if the mentioned
inverses exist). The first four
follow inmmediately. The fifth one follows from
\begin{eqnarray}
B^TA^{-1\,T}&=&B^TA^{-1\,T} (A^TD-C^TB)= B^TD-B^TCA^{-1}B\nonumber\\
&=& B^T D+ (\unity -D^TA)A^{-1}B=A^{-1}B \label{Am1Bsym} \ ,
\end{eqnarray}
after which the others are easy too.
\begin{lemma}          \label{lemma553}
If the $2m\times m$ matrix $V=\pmatrix {X^I\cr Y_I}$, where
$X^I$ and $Y_I$ are $m$-vectors, is of rank $m$, then there
exists a symplectic rotation ${\cal S}\in \symp{2m}$ such that
$\tilde V={\cal S}V=\pmatrix {\tilde X\cr \tilde Y}$ has the
property that $\tilde X$ is invertible.
\end{lemma}
\begin{proof}
We introduce the notation
\begin{equation}
X^r \sim \left[  \rvec{X}{1}{r-1} \right]
\end{equation}
to indicate that $X^r$ is a linear combination of the vectors
$X^1,\ldots X^{r-1}$, and
\begin{equation}
0 \not\sim [\rvec{X}{1}{r-1}]
\end{equation}
means that these vectors are independent.

We will explicitly construct the matrix ${\cal S}$. Take the
following hypothesis of induction.
\begin{eqnarray} \label{assume1}
0 &\not\sim& [\rvec{X}{1}{r-1},Y_k]\ ;\\
\label{assume2}
X^r & \sim & \left[ \rvec{X}{1}{r-1} \right]\ .
\end{eqnarray}
So $V$ is already of rank $r-1$, and assuming $r-1<m$, some vector
$Y_k$ should be independent of $\rvec X1{r-1}$. We now
symplectically rotate this vector into the $X$ part. The following cases are
distinguished:

\noindent{\bf Case 1}: \quad $k=r$. \newline
Consider the symplectic matrix
\begin{equation}
S_r = \left( \begin{array}{cc}
       \unity-E_{r,r} & E_{r,r} \\
       -E_{r,r} & \unity -E_{r,r}
           \end{array} \right)   \ ,\label{Sr}
\end{equation}
i.e.
\begin{eqnarray*}
  \tilde{X}^I &=& X^I \ ,\qquad I \not= r \ ;\\
  \tilde{X}^r &=& Y_k  \ .
\end{eqnarray*}
This symplectic rotation takes the matrix into one of which
the first r rows are independent.
\par
\noindent{\bf Case 2}: \quad $k\not=r$. \newline
In this case we use the symplectic matrix
\begin{equation}
 S_{k,r} = \left(
\begin{array}{cc} \unity-E_{r,r}-\frac{1}{\alpha}E_{k,r} & E_{r,k} + \alpha E_{r,r} \\
                  -\frac{1}{\alpha} E_{r,r} & \unity - E_{r,r} \\
\end{array} \right)    \ .
\label{Skr}\end{equation}
In components this reads
\begin{eqnarray}
  \nonumber
  \tilde{X}^{I} &=& X^I \qquad \quad I \not= k,r\ ;\\ \nonumber
  \tilde{X}^k &=& X^k-\frac{1}{\alpha}X^r \ ;\\
\label{newXR}
  \tilde{X}^r &=& Y_k+\alpha Y_r \ ;\\            \nonumber
  \tilde{Y}_I &=& Y_I \qquad \quad  I \not=  r\ ;\\  \nonumber
  \tilde{Y}_r &=& -\frac{1}{\alpha}X^r \ .
\end{eqnarray}
Notice the introduction of the parameter $\alpha$ whose value can be taken
to be arbitrary at this stage. Its relevance will be discussed below.

Next we find out about the conditions which will guarantee that
$\tilde{X}^1,\ldots,\tilde{X}^r$ are independent m-vectors. Let
\begin{equation}
\label{indepeq} \sum^{r}_{I=1}\alpha_{I}\tilde{X}^I = 0\ .
\end{equation}
Substituting the transformation rules (\ref{newXR})
gives
\begin{eqnarray}\label{step1}
\sum^{r-1}_{I=1,I\not= k}\alpha_{I}X^I
+ \alpha_{k}\big(X^k-\frac{1}{\alpha}X^r \big)
+ \alpha_{r}\big( Y_k + \alpha Y_{r}\big)= 0\ ,
\end{eqnarray}
with the convention $\alpha_{k} = 0$ for $k > r$. The hypothesis of induction
\eqn{assume2}  states that
\begin{equation}
X^r = \sum_{J=1,J\not= k}^{r-1}\lambda_{J} X^J +
\mu X^k \ ,
\end{equation}
for some $\lambda_{J}, \mu $ with once more the convention
$\mu = 0$ for $k > r$. Upon substitution into
(\ref{step1}) one finally finds
\begin{equation} \label{newindepeq}
\sum^{r-1}_{I=1,I\not= k}
(\alpha_{I}-\lambda_{I}\frac{\alpha_{k}}{\alpha}) X^I
+ \alpha_{k}(1-\frac{\mu}{\alpha}) X^k+ \alpha_{r} Y_{k}
+ \alpha_{r}\alpha Y_r = 0\ .
\end{equation}
The following cases are distinguished:
\begin{enumerate}
\item $Y_r \not\sim [X^1,\ldots X^{r-1},Y_k]\ . $ \\
For $\alpha_{k}$ different from zero a priori (i.e. $k < r$) it suffices
to pick $\alpha \not= \mu$ in order that all $\alpha_{I}$ vanish for $I =
1,\ldots,r$. This is an immediate corollary from the assumed independence
of the set $X^{I}, Y_{r}$ (for $I=1\ldots,r-1$) and $Y_k$ in (\ref{newindepeq}).
\item $Y_r \sim [X^1,\ldots X^{r-1},Y_k]\ . $ \\
Equivalently,
\begin{equation}
Y_{r} = \sum^{r-1}_{I=1,I\not= k} \beta_{I} X^I +
\beta_{k} X^k + \nu Y_k\ ,
\end{equation} for appropriate values of the coefficients. In this way
(\ref{newindepeq}) reads alternatively
\[
\sum^{r-1}_{I=1,I\not= k} (\alpha_{I}-\lambda_{I}\frac{\alpha_{k}}{\alpha}
+ \beta_{I}\alpha_{r}\alpha) X^I +
(\alpha_{k}(1-\frac{\mu}{\alpha})+
\alpha_{r}\alpha\beta_{k})X^k
\]
\begin{equation}
+ \alpha_{r}(1 + \nu\alpha) Y_k = 0 \ .
\end{equation}
All $X^{I}$ and $Y_k$ occuring in this equation are once more assumed to be independent
quantities, implying the vanishing of the coefficients in front.
Explicitly,
\begin{eqnarray}
(\alpha_{I}-\lambda_{I}\frac{\alpha_{k}}{\alpha} + \beta_{I}\alpha_{r}\alpha)&=& 0
\\ \nonumber
\quad \mbox{ for } \quad I = 1,\ldots,r-1;I \not= k\ ; \\ \label{cfXk}
(\alpha_{k}(1-\frac{\mu}{\alpha})
+ \alpha_{r}\alpha\beta_{k}) &=& 0 \ ; \\
 \label{cfFk}
\alpha_{r}(1 + \nu\alpha) &=& 0 \ .
\end{eqnarray}
It is here that the relevance of the freedom to pick an arbitrary value
for $\alpha $ emerges. For if we can choose $\alpha $ such that $1 + \alpha\nu \not= 0$,
then (\ref{cfFk}) implies $\alpha_{r} = 0$. By the same token one
concludes that the ability to choose $\alpha \not= \mu$ ( in the case
$k<r$ ) forces $\alpha _{k}$ to vanish by (\ref{cfXk}). Furthermore, all
remaining $\alpha_{I}$ in the first equation become identically zero.
\end{enumerate}
We conclude that with the hypothesis of induction (\ref{assume1}) and
(\ref{assume2}) the following implication holds:
\begin{equation}
\sum^{r}_{I=1}\alpha_{I}\big(\tilde{X}^I\big) = 0 \Rightarrow \forall I:\
\alpha_{I} = 0\ .
\end{equation}
This simply states the independence of the quantities
$\tilde{X}^1,\ldots,\tilde{X}^r$.

Some small comment is appropriate. In the above reasoning it was assumed that
the constant parameter $\alpha $ could be chosen appropriately. At first
sight this might conflict with the fact that the m-vectors $X^I$
 (as well as the $Y_I$) are in general functions of $z^\alpha$ and
$\bar z^{\bar \alpha}$. Recall the restrictions on $\alpha$~:
\begin{eqnarray} \label{cona1}
\alpha &\not=& \mu\ ; \\  \label{cona2}
\alpha &\not=& -\frac{1}{\nu}\ ;
\end{eqnarray}
These could possibly constrain the domain in which the above procedure is
valid. Note however, that if the m-vectors depend continuously on $z^\alpha$
and $\bar z^{\bar \alpha}$ one can always find some value for $\alpha $
such that the constraints (\ref{cona1}) and (\ref{cona2}) are obeyed
on some sufficiently small compact region $W \subset {\cal M}$. In any
case some neighbourhood on which the algorithm is a valid procedure, is
contained within this compact region.
\end{proof}

\begin{property}
   \label{prop:sympfact}
Every \symp{2m}-matrix can be written as the product of a finite
number of symplectic matrices of the following form:
\begin{equation}
L=\pmatrix{ A&0\cr C&{A^{-1T}} }\ ;\qquad
S=\pmatrix{0&\unity\cr -\unity&0} \ .
\end{equation}
The property of symplecticity implies that $A^TC$ is symmetric.
\end{property}

\begin{proof}
If  $A$ is invertible, the statement is proved by the decomposition
\begin{equation}
\matrx ABCD = \matrx{A}{0}{C}{A^{-1T}}\matrx 0\unity {-\unity}0
\matrx{-\unity}{0}{A^{-1}B}{-\unity}\matrx 0\unity {-\unity}0\ .
\end{equation}
The only non-trivial part of this equation is
\begin{equation}
CA^{-1}B+A^{-1T} =A^{-1T}(C^TB+\unity )=A^{-1T}A^TD=D\ .
\end{equation}
If $A$ is not invertible, the last of \eqn{symplABCD} still implies
that \[ \left(\begin{array}{c}A\\C \end{array}
\right)\]
has maximal rank. Therefore we can use lemma~\ref{lemma553} to
transform it to the case of an invertible $A$. Reviewing the proof of
the latter lemma, we see that we performed symplectic rotations of
the form \eqn{Sr} or \eqn{Skr}. Both of them can be written as the
product of symplectic matrices with an invertible $A$-matrix, e.g.
for $\alpha\neq 1$:
\begin{equation}
S_{k,r}= \matrx{\unity}{\unity}{0}{\unity}
\left(
\begin{array}{cc} \unity+(\ft1\alpha-1)E_{r,r}-\ft{1}{\alpha}E_{k,r} &
E_{r,k}-\unity +(1+\alpha) E_{r,r} \\
                  -\ft{1}{\alpha} E_{r,r} & \unity- E_{r,r}
\end{array} \right) \ .
\end{equation}
Therefore all symplectic matrices are written as products of matrices
with an invertible $A$-part, which, using the first part of this
proof, implies the statement.
\end{proof}
\pagebreak[2]

\begin{lemma}   \label{N}
A symmetric matrix \N\ with negative-definite imaginary
part is invertible.
\end{lemma}          \nopagebreak[4]
\begin{proof}
The following identities hold for an arbitrary complex
vector  $z=x+iy$:
\begin{eqnarray*}
 z^\dagger\N z & = & (x^T-iy^T)\,\N\,(x+iy)\\
         & = & x^T\N x+y^T\N y-iy^T\N x+ix^T\N y  \ .
 \end{eqnarray*}
 Because of the symmetry of \N\ the last two terms cancel.
 So
\[\IM ( z^\dagger\N z )=x^T(\IM\N)x+y^T(\IM\N)y\ ,\]
from which it follows that each nullvector $z$ of \N\
equals zero.
\end{proof}
\begin{theorem}  \label{pq}
Take $\left( \begin{array}{cc}
A & B \\ C & D
        \end{array}       \right) \in \Sp{2m}$ . Consider $(m\times
        m)$-matrices $X$ and $Y$ such that
 \begin{enumerate}
  \item $X$ is invertible; \label{inv}
  \item $YX^{-1} $ is symmetric;  \label{symm}
  \item $\IM (YX^{-1})<0 \ .$          \label{im}
\end{enumerate}
Define $\left( \begin{array}{c}
                  \tilde{X} \\ \tilde{Y}
                  \end{array}   \right)
 = \left( \begin{array}{cc}
         A & B \\ C & D
        \end{array}       \right)
    \left( \begin{array}{c}
                  X \\
                  Y
                  \end{array}   \right)$,
then $\tilde{X}$ and $\tilde{Y}$ satisfy the conditions~\ref{inv}
till~\ref{im}.
\end{theorem}

\begin{proof}
Note that proposition  \ref{prop:sympfact}  implies that we
need only prove the theorem for the following cases:
\begin{enumerate}
\item $\left( \begin{array}{cc}
              A & B \\ C & D
              \end{array}       \right)
       =\left( \begin{array}{cc}
              0 & \unity \\ -\unity & 0
              \end{array}       \right)\ .$
\item $\left( \begin{array}{cc}
A & B \\ C & D
              \end{array}       \right)
       =\left( \begin{array}{cc}
              A & 0 \\ C & D
              \end{array}       \right)\ .$
\end{enumerate}
\begin{itemize}
\item
First we show that  $\tilde{X}$ is invertible in both
cases:
 \begin{enumerate}
    \item $\tilde{X}=Y$. This is invertible because the
    items \ref{symm}
    and
    \ref{im} and lemma~\ref{N} imply that $YX^{-1}$ is.
    \item $\tilde{X}=AX$ is invertible since $A$ and $X$ are.
    \end{enumerate}
\item $\tilde{Y}\tilde{X}^{-1}$ is symmetric:
\[\tilde{Y}\tilde{X}^{-1}-\tilde{X}^{-1T}\tilde{Y}^{-1T}
=\tilde{X}^{-1T}(\tilde{X}^T\tilde{Y}-\tilde{Y}^T\tilde{X})\tilde{X}^{-1}\ .\]
Because of symplecticity the expression in brackets equals $(X^T Y-Y^T X)$, which is
zero since $YX^{-1}$ is symmetric.
\item $\IM (\tilde{Y}\tilde{X}^{-1})<0$: \newline
{}From the symmetry of $ \tilde{Y}\tilde{X}^{-1} $ we get
$( \tilde{Y}\tilde{X}^{-1})^{\star}
=(\tilde{Y}\tilde{X}^{-1} )^\dagger$, so
\begin{eqnarray*}
 \IM (\tilde{Y}\tilde{X}^{-1})&=&\frac{1}{2i}[ \tilde{Y}\tilde{X}^{-1} -
 (\tilde{Y}\tilde{X}^{-1})^\dagger]\\
 &=&\frac{1}{2i} \{\tilde{X}^{-1+}[\tilde{X}^\dagger\tilde{Y}-\tilde{Y}^\dagger\tilde{X}]
 \tilde{X}^{-1}\}     \ .
\end{eqnarray*}
The expression in brackets is the symplectic inner product of
$(\tilde X^*,\tilde Y^*)$  and $(\tilde X,\tilde Y)$  and thus equals
$[X^\dagger Y-Y^\dagger X]$, so
\[\IM ( \tilde{Y}\tilde{X}^{-1} )=(X\tilde{X}^{-1})^\dagger \IM (YX^{-1})( X\tilde{X}^{-1})\ ,\]
which is negative-definite because of property~\ref{im}
and the invertibility of
$  X\tilde{X}^{-1}$.
\end{itemize}
\end{proof}

\begin{gevolg}  \label{gevolg}
If \N\ is a symmetric ($m \times m$)-matrix with
strictly negative imaginary part
and $\left( \begin{array}{cc}
         A & B \\ C & D
        \end{array}       \right) \in \Sp{2m}$, then
$A+B\N$ is invertible. Furthermore,
$(C+D\N)(A+B\N)^{-1}$ is also a symmetric matrix with strictly negative
imaginary part.
\end{gevolg}

\begin{proof}
 Take in proposition~(\ref{pq}) $X=\unity$ and $Y=\N$.
\end{proof}

\section{Theorems on special geometry \label{1=2lemmas}}
\begin{lemma}          \label{lemma551}
If $V$ satisfies the conditions of remark~1,
page~\pageref{remark1}, and
$g_{\alpha\bar\beta}$ is non-degenerate then
\[
\cal{W}\equiv\left(
      \begin{array}{cc}
         X^I&F_I\\\Dee{\alpha}X^I&\Dee{\alpha}F_I
      \end{array} \right)
\]
(where $I=\rng{0}{n}$ and $ \alpha=\rng{1}{n}$) has rank $n+1$.
\end{lemma}
\begin{proof}
Suppose ${\rm rank}(\cal{W}) \leq$ n,
then there exist $\lambda^{\alpha}$ and $\lambda^{0}$, not all zero,
such that
\begin{equation}
 \lambda^{\alpha}U_{\alpha}+\lambda^{0} V  =0\ .
\end{equation}
Taking first an inner product with $\bar V$, and using \eqn{UbarV0}
and condition~\ref{eisVVbar}, gives $\lambda^0=0$, and thus
$\lambda^\alpha$ is not trivial. Then taking the inner product with
$\bar U_{\bar\beta}$, and using \eqn{UbarV0} and \eqn{UbarUg}, we get
$ \lambda^{\alpha}g_{\alpha\bar\beta}=0$, which is in contradiction
with our assumption that $g_{\alpha\bar\beta}$ is non-degenerate.
\end{proof}
\pagebreak[2]

\begin{lemma} \label{lemma554}
\[\det \mtrx{\Dee{\alpha}X^B}{\Dee{\alpha}X^0}{X^B}{X^0}\ne 0 \Leftrightarrow
\det \left[\dee{\alpha}\left(\frac{X^B}{X^0}\right)\right]\ne 0\]
if $X^0\ne 0$.
\end{lemma}
\begin{proof}
The proof follows from the equality of the following determinants:
\begin{eqnarray}
\lefteqn{\det \mtrx{\Dee{\alpha}X^B}{\Dee{\alpha}X^0}{X^B}{X^0}=
\det \mtrx{\dee{\alpha}X^B}{\dee{\alpha}X^0}{X^B}{X^0}}\nonumber\\
& =& \det \mtrx{\dee{\alpha}X^B-
\frac{X^B}{X^0}\dee{\alpha}X^0}{\dee{\alpha}X^0}{X^B-X^B}{X^0}=
(X^0)^{n+1}\det \left[\dee{\alpha}\left(\frac{X^B}{X^0}\right)\right]
\ .
\end{eqnarray}
\end{proof}

\begin{lemma}  \label{lemma555}
If\/ $\det \pmatrix{\Dee{\alpha}X^I\cr X^I}\ne 0$
and $\Dee{\bar\alpha}\left( \begin{array}{c}
 X^I\\F_I
 \end{array}\right) =0$, then some functions
$ \tilde F_I(X^J)$ exist such that
\begin{equation}
\label{FvanXvanz}
 \tilde F_I(X^J(\za,\bar z^\alpha ))=F_I(\za ,\bar z^\alpha) \ ,
\end{equation}
and
\begin{equation}  \label{FtildeI}
 \tilde F_I=\XJ\dee{J}\tilde F_I\ .
\end{equation}
\end{lemma}

\begin{proof}
Because of $\Dee{\bar\alpha}\left( \begin{array}{c}
 X^I\\F_I
 \end{array}\right) =0$, $\frac{X^B}{X^0}$ and $\frac{F_I}{X^0} $
   are holomorphic functions of the
 \za. Lemma~\ref{lemma554} guarantees that one can express the \za\ in terms of the
 independent variables
 $\frac{X^B}{X^0}$. Now set e.g.
\begin{equation}
\tilde F_I(X^J)\equiv X^0f_I\left(\za\left(\frac{X^B}{X^0} \right)\right)\ ,
\end{equation}
with
\begin{equation}
f_I(\za)\equiv \frac{F_I(\za,\bar z^\alpha )}{X^0 (\za,\bar z^\alpha )}\ .
\end{equation}
Equation \eqn{FtildeI} is immediate for $\tilde F_I$ is homogeneous of
first degree.
\end{proof}
\begin{lemma}  \label{lemma556}
If\/ $\det \pmatrix{\Dee{\alpha}X^I\cr X^I}\ne 0$,
conditions~\ref{eisUaUb} and~\ref{eisVUa} of remark~1
(page~\pageref{remark1}) hold, and \eqn{FvanXvanz} and \eqn{FtildeI}
hold, then some function $\tilde F(X^J)$ exists such that
\begin{equation}
  \tilde F_I=\dee{I}\tilde F(X^I)\ .
\end{equation}
\end{lemma}

\begin{proof}
The two conditions of remark~1 can, with the mentioned equations of
lemma~\ref{lemma555}, be written respectively as
\begin{eqnarray}
0&=&\Dee{\alpha}X^I\Dee{\beta}X^J\dee{\left[I\right.}
\tilde F_{\left. J\right]}\nonumber\\
0&=&X^I\dee{\alpha}F_I-F_I\dee{\alpha}X^I
=\dee{\alpha}X^I X^J\dee{\left[I\right.}\tilde F_{\left. J\right]}
=\Dee{\alpha}X^I X^J\dee{\left[I\right.}\tilde F_{\left. J\right]} \ .
\end{eqnarray}
Because of the invertibility of
$\left(\begin{array}{c} \Dee{\alpha}X^I \\ X^I \end{array}\right)$,
these equations imply that
$\dee{\left[I\right.}\tilde F_{\left. J\right]}=0$, which is the
integrability condition for the existence of $\tilde F(X)$.
\end{proof}
\begin{lemma}
If the matrix
\begin{equation}
i\pmatrix{\langle U_\alpha ,\bar U_{\bar \beta}\rangle &
\langle U_\alpha ,V\rangle
\cr \langle \bar V,\bar U_{\bar \beta}\rangle&\langle \bar V,V\rangle }
\label{matreqcN}\end{equation}
is positive definite, then the matrix
$\left( \begin{array}{cc} \Dee{\alpha} X^I&\bar X^I
\end{array}\right)$
is invertible.  \label{lem:invbarVU}
\end{lemma}
\begin{proof}
Suppose a linear combination of the columns of  this matrix equals zero:
\begin{equation}
a^\alpha\Dee{\alpha}X^I+b\bar X^I=0                    \ .
\end{equation}
Then the same statement applies to the complex conjugate equation. So it
follows immediately that
\begin{equation}
\langle a^\alpha U_\alpha+b\bar V,\bar a^{\bar\beta}\bar
U_{\bar\beta}+\bar b V\rangle=0 \ .
\end{equation}
This can be written as follows:
\begin{equation}
-i\pmatrix{a^\alpha&b} \pmatrix{\langle U_\alpha ,\bar U_{\bar \beta}\rangle &
\langle U_\alpha ,V\rangle
\cr \langle \bar V,\bar U_{\bar \beta}\rangle&\langle \bar V,V\rangle }
\pmatrix{\bar a^{\bar\beta}\cr \bar b} =0  \ .
\end{equation}
The positivity of \eqn{matreqcN} implies that $a=b^\alpha=0$.
\end{proof}
\begin{gevolg}  \label{cor:gevolg2}
If the conditions $\langle V,\bar V\rangle =i$ and
$\langle V,U_{\alpha}\rangle =0$
are satisfied and
the metric
$g_{\alpha\bar \beta}\equiv i\langle U_\alpha,\bar U_{\bar\beta}\rangle$
is positive definite then the matrix
$\left( \begin{array}{cc} \Dee{\alpha} X^I&\bar X^I
\end{array}\right)$
is invertible.
\end{gevolg}

\section{Counterexamples}   \label{counterexample}
\subsection{Example 1}
\label{app:example1}
We give an explicit example of a \Ka\ manifold which
satisfies Strominger's definition of a special \Ka\
manifold, but which is not special according to our
definitions. This example proves the non-triviality of the
condition \eqn{cond2local} for $n=1$.

Our manifold is the following part of the complex plane: an
open disk centered at 0 with sufficiently small radius (cf.
infra). The \Ka\ metric is obtained from the \Ka\ potential
\begin{equation}
\label{vb:K}
K=-\log \left(|1+z^4+z^5+z^6+z^7|^2-|z|^2 +|z|^4-|z|^6 \right)\ .
\end{equation}
The disk's radius should be small enough for the metric to
be positive definite. Since  $g_{z{\bar z}}(0)=1$ a
suitable radius exists.

First we prove that  $g_{z{\bar z}} $ is special according
to Strominger's definition, i.e. that there exists a
holomorphic symplectic four-component vector $v(z)$ such
that
\begin{equation}
\label{vb:e-K}
e^{-K}=i\langle\bar v,v\rangle\ .
\end{equation}

\begin{proof}
Note that
\begin{equation}
\label{vb:hermvorm}
(x,y)\mapsto i\langle\bar x,y\rangle\ ;\qquad x,y\in \Cbar^4
\end{equation}
defines a non-degenerate sesquilinear form. The following
vectors constitute a basis of $ \Cbar^4 $ which is
orthonormal with respect to this form:

\begin{equation}
v_0=\left(\begin{array}{c} 1\\0\\ -\frac{i}{2}\\0\end{array}
\right)                     \ ;\qquad
v_1=\left(\begin{array}{c} 1\\0\\ \frac{i}{2}\\0\end{array}
\right)                       \ ;\qquad
v_2=\left(\begin{array}{c} 0\\1\\0\\ -\frac{i}{2} \end{array}
\right)                        \ ;\qquad
v_3=\left(\begin{array}{c} 0\\1\\0\\ \frac{i}{2}\end{array}
\right)                       \ .
\end{equation}
Indeed:
\begin{equation}
\label{vb:ONB}
\langle v_k,\bar v_l\rangle=i\delta_{k,l}(-1)^k \ .
\end{equation}
Now define
\begin{equation}
\label{vb:vormv}
v\equiv v_0(1+z^4+z^5+z^6+z^7)+v_1z+v_2z^2+v_3z^3  \ .
\end{equation}
Then it follows from~\eqn{vb:ONB} that
\begin{equation}
\langle v,\bar v\rangle=i(1+z^4+z^5+z^6+z^7)
(1+\bar z^4+\bar z^5+\bar z^6+\bar z^7)
-iz{\bar z}+iz^2{\bar z}^2-iz^3{\bar z}^3\ ,
\end{equation}
and thus \eqn{vb:e-K}.
\end{proof}\vspace{3mm}

Consider the point $z=0$, which is in our domain. There $V=v=v_0$ and
$U_z=\partial_z v=v_1$. So we have $\sinprod {U_\alpha}V=\sinprod
{\partial_z v}v=-i$, so this
violates \eqn{cond2local} or the condition~\ref{eisVUa},
page~\pageref{eisVUa}. The matrix \eqn{ImcN} is at this point
\begin{equation}
\left.i\pmatrix{\langle U_z ,\bar U_{\bar z}\rangle &
\langle U_z ,V\rangle
\cr \langle \bar V,\bar U_{\bar z}\rangle&\langle \bar V,V\rangle }
\right|_{z=0}= \pmatrix{1&1\cr 1&1}\ ,
\end{equation}
which is thus not invertible. Also \eqn{invbarVU} is then not
invertible, and we have no unique definition of ${\cal N}$. Therefore
this symplectic section cannot be used for supergravity.

Finally we prove that  $g_{z{\bar z}}$ is not special
according to our second definition, i.e. that no
holomorphic symplectic vector $w(z)$ exists which satisfies
both
\begin{equation}              \label{vb:e-Kbis}
 e^{-K}=i\langle\bar w,w\rangle
\end{equation}
and
\begin{equation}             \label{vb:vdv}
\langle w,\dee{z}w\rangle=0  \ .
\end{equation}
[A priori  we should impose instead of \eqn{vb:e-Kbis}
the weaker condition
\begin{equation}
\label{vb:verzw}
\exists f(z):\qquad  i\langle\bar w,w\rangle =e^{-K+f(z)+\bar f({\bar
z})}\ ,
\end{equation}
but then  $e^{-f(z)}w$ would satisfy \eqn{vb:e-Kbis} and
\eqn{vb:vdv} whenever
$w$ satisfied \eqn{vb:verzw} and \eqn{vb:vdv}.
So what we will prove is sufficiently general.]

\begin{proof}
Suppose there exists a holomorphic vector $w$ satisfying
equations \eqn{vb:e-Kbis}
and \eqn{vb:vdv}. We show that this leads to a contradiction.

We can write $w$ as a power series:
\begin{equation}
w=w_0+w_1z+w_2z^2+\ldots
\end{equation}
with $w_k\in \Cbar^4$ as yet arbitrary coefficients.
Then
\begin{equation}
\langle w,\bar w\rangle=\sum_{k,l} \langle w_k,\bar w_l\rangle z_k{\bar
z}_l \ .
\end{equation}
Identifying the coefficients
$ \langle w_k,\bar w_l\rangle $ and those of
$ie^{-K}$,  we find
\begin{eqnarray}
\langle w_k,\bar w_l\rangle= i\delta_{k,l}(-1)^k&\mbox{voor}&k,l=0,1,2,3\
;\label{vb:ONBbis}\\
\langle w_k,\bar w_0\rangle= i  &\mbox{voor}&4\leq k\leq 7\ ;\label{vb:vkv0bar}\\
\langle w_k,\bar w_l\rangle= 0  &\mbox{voor}&4\leq k\leq 7\ ,\qquad l=1,2,3\ .
\label{vb:vkvlbaris0}
\end{eqnarray}
Since $\{w_0,w_1,w_2,w_3\}$ is an orthonormal basis (see~\eqn{vb:ONBbis}),
equation \eqn{vb:vkvlbaris0} implies that $w_k$ is proportional to
$w_0$ for $4\leq k\leq 7$.
Then it follows from equations \eqn{vb:vkv0bar} and \eqn{vb:ONBbis} that
\begin{equation}  \label{vb:vkisv0}
w_k=w_0\ ,\qquad 4\leq k\leq 7 \ .
\end{equation}
Summing up, equation \eqn{vb:e-Kbis} implies that $\{w_0,w_1,w_2,w_3\}$ is an
orthonormal basis  (with respect to \eqn{vb:hermvorm}) and that $w_k=w_0 $ for
$4\leq k\leq 7$.

Now we write out equation \eqn{vb:vdv}:
\begin{equation}
0=\langle w,\dee{z}w\rangle =
\langle w_0+w_1z+w_2z^2+\ldots,w_1+2w_2z+3w_3z^2+\ldots\rangle   \ .
\end{equation}
All terms of the power series should vanish.
We put the coefficients of  $z^0$, $z^1$ and $z^6$ equal to
zero and take equation  \eqn{vb:vkisv0}  and the
antisymmetry of $\langle.,.\rangle$ into account:
\begin{eqnarray*}
\langle w_0,w_1\rangle&=&0\ ;\\
\langle w_0,w_2\rangle&=&0\ ;\\
\langle w_3,w_0\rangle&=&0   \ .
\end{eqnarray*}
These express that $\bar w_0$ is orthogonal  to
$w_1$, $w_2$ and $w_3$. Furthermore,  $\bar w_0$ is clearly
orthogonal to
$w_0$ ($\langle w_0,w_0\rangle=0 $).
Thus is
\begin{equation}
\bar w_0=0\ .
\end{equation}
This is in contradiction with $\langle w_0,\bar w_0\rangle=i$.
\end{proof}

\subsection{Example 2}
\label{app:example2}
As a second example, consider
\begin{equation}
V=e^{K/2}\pmatrix{1\cr z\cr -iz\cr -ia \cr}\ ,
\end{equation}
with $a\in\Rbar$. It satisfies conditions~\ref{eisVVbar}
and~\ref{eisDabarV} of remark~1. For $a=1$ this is the symplectic
section~\ref{examplev}, originating from
the prepotential $F=-i X^0 X^1$, and for all $a$ it has the same \Ka\
potential up to a \Ka\ transformation:
\begin{equation}
e^{-K}= (1+a)(z+\bar z)\ ;\qquad\partial_z\partial_{{\bar z}}K=
(z+\bar z)^{-2}\ ,
\end{equation}
with a positivity domain if  $a\neq -1$, for example $a>-1$ and $\Re z>0$.
We find
\begin{equation}
U_z=\frac{1}{z+\bar z}e^{K/2}\pmatrix{-1\cr \bar z\cr -i\bar z\cr ia \cr}
\end{equation}
and $\langle U_z, V\rangle =i(1-a)e^K$.
So the constraint \eqn{cond2local} selects $a=1$.
However, the matrix \eqn{ImcN},
\begin{equation}
i\pmatrix{\langle U_z ,\bar U_{\bar z}\rangle &
\langle U_z ,V\rangle
\cr \langle \bar V,\bar U_{\bar z}\rangle&\langle \bar V,V\rangle }
= \pmatrix{(z+\bar z)^{-2}&\frac{a-1}{a+1}(z+\bar z)^{-1}\cr
\frac{a-1}{a+1}(z+\bar z)^{-1}&1}
\end{equation}
remains positive definite for all $a>0$. Therefore we can define the
matrix ${\cal N}$, which is
\begin{equation}
{\cal N}=\pmatrix{-i z&0\cr 0&-\frac{i a}{ z}}\ ,
\end{equation}
i.e. symmetric and negative definite imaginary part for $a>0$, as
follows from the general arguments in section~\ref{ss:def2}.
In the present constructions of $N=2$ supergravity couplings this matrix
for $a\neq 1$ is never obtained because it violates
\begin{equation}
-\ft12(\Im {\cal N})^{-1}=\bar f_{\bar \alpha}^I g^{\bar \alpha\beta}
f_\beta^J + X^I\bar X^J \ .   \label{reqcN}
\end{equation}
In fact, this relation is a rewriting of \eqn{ImcN}, using in the
left hand side the usual constraints of special geometry, including
$\sinprod {U_\alpha}V=0$. This equation is used in proving the
supersymmetry commutator on the vector field, or the invariance of
the Pauli-terms in the action, where the two terms on the right hand
side correspond to contributions via the spin-$\frac{1}{2}$ and
spin-$\frac{3}{2}$ fermions respectively. However, we know no
physical argument to exclude the possibility of a modification in
case $\sinprod {U_\alpha}V\neq 0$.


\newpage

\begin{thebibliography}{99}
\bibitem{DWVP}  B. de Wit, P.G. Lauwers, R. Philippe, Su S.-Q.
and A. Van Proeyen, \Journal{\PLB}{134}{37}{1984};\\
B. de Wit and A. Van Proeyen, \Journal{\NPB}{245}{89}{1984}.
\bibitem{special} A.~Strominger, \Journal{\CMP}{133}{163}{1990}.
\bibitem{PKTN2} G. Sierra and P.K. Townsend, in {\em
Supersymmetry and Supergravity 1983}, ed. B. Milewski (World
Scientific, Singapore, 1983);\\
S. J. Gates, \Journal{\NPB}{238}{349}{1984}.
\bibitem{CdAF}
L. Castellani, R. D' Auria and S. Ferrara,
\Journal\PLB {241}{57}{1990};
\Journal\CQG{7}{1767}{1990};
\\
R.~D'Auria, S.~Ferrara and P.~Fr\`e, \Journal\NPB{359}{705}{1991}.
\bibitem{CDFLL}
S. Ferrara and J. Louis, \Journal\PLB{278} {240}{1992};\\
A. Ceresole, R. D'Auria, S. Ferrara, W. Lerche and J. Louis,
\Journal\IJMPA 8{79}{1993} , hep-th/9204035; \\
J. Louis, in {\em String theory and quantum gravity '92},
eds. J. Harvey et al. (World Scientific, Singapore,1992), p. 368.
\bibitem{f0art} A. Ceresole, R. D'Auria, S. Ferrara and A. Van
Proeyen, \Journal\NPB {444} {92}{1995}, hep-th/9502072.
\bibitem{prtrquat} B. de Wit and A. Van Proeyen,
Proceedings of the Meeting on
 Quaternionic Structures in Mathematics and Physics, Trieste,
 September 1994, ILAS/FM-6/1996, p.109; hep-th/9505097.
\bibitem{fresoriabook} P. Fr\`e and P. Soriani, {\em The $N=2$
Wonderland, from Calabi--Yau manifolds to topological
field--theories}, (World Scientific, Singapore, 1995).
\bibitem{ItalianN2}
L. Andrianopoli, M. Bertolini, A. Ceresole, R. D'Auria,
S. Ferrara and P. Fr\'e, Nucl. Phys. {\bf B476} (1996)
397; hep-th/9603004\\
L. Andrianopoli, M. Bertolini, A. Ceresole, R. D'Auria,
S. Ferrara, P. Fr\'e and T. Magri, hep-th/9605032.
\bibitem{sc2defsg}
B. Craps, F. Roose, W. Troost and A. Van Proeyen, preprint
KUL-TF-96/11, hep-th/9606073, to be published in the proceedings
of the second
international Sakharov conference on physics, Moscow, May 1996.
\bibitem{benfredlic}
B. Craps and F. Roose, licentiaatsthesis, K.U. Leuven, 1996.
\bibitem{dual} S. Ferrara, J. Scherk and B. Zumino,
\Journal\NPB{121}{393}{1977};\\
B. de Wit,  \Journal\NPB{158}{189}{1979}; \\
E. Cremmer and B. Julia, \Journal\NPB {159}{141}{1979};\\
M.K. Gaillard and B. Zumino, \Journal\NPB {193}{221}{1981}.
\bibitem{CecFerGir} S. Cecotti, S. Ferrara and L. Girardello,\Journal\IJMPA
4 {2457}{1989}.
\bibitem{UCCVarDim} L. Andrianopoli, R. D'Auria and S. Ferrara,
hep-th/9612105.
\bibitem{N2prepotentials} L. Mezincescu, JINR preprint P2 (1979)
1257;\\
J. Koller, Nucl. Phys. {\bf B222} (1983) 316; Phys. Lett. {\bf B124}
(1983) 319;\\
P. Howe, K. Stelle and P. Townsend, Nucl. Phys. {\bf B236} (1984)
125.
\bibitem{trsummer} A. Van Proeyen,
in  {\it
1995 Summer school in High Energy Physics and Cosmology}, The ICTP
series in theoretical physics - vol.12 (World Scientific, 1997), eds.
E. Gava et al., p.256;
 hep-th/9512139.
\bibitem{vectorm} P. Fayet, Nucl. Phys. {\bf B113} (1976) 135;
{\bf B149} (1979) 137;\\
R. Grimm, M. Sohnius and J. Wess, Nucl. Phys. {\bf B133} (1978) 275.
\bibitem{FI} P. Fayet and J. Iliopoulos, \Journal\PLB{51}{461}{1974}
\bibitem{APT} I. Antoniadis, H. Partouche and T.R. Taylor, \Journal\PLB
{372}{83}{1996}, hep-th/9512006; \\
I. Antoniadis and T.R. Taylor,
Fortsch.Phys. {\bf 44} (1996) 487; hep-th/9604062.
\bibitem{BEC} E. Cremmer, C. Kounnas, A. Van Proeyen, J.P. Derendinger, S.
Ferrara, B. de Wit and L. Girardello, \Journal\NPB {250} {385}{1985}.
\bibitem{hidden} B. de Wit and A. Van Proeyen,
Phys. Lett. {\bf B252}  (1990) 221;
{\bf B293} (1992) 94; Commun. Math. Phys.
{\bf 149}, 307 (1992);\\
B.~de~Wit, F.~Vanderseypen and A.~Van~Proeyen,
Nucl. Phys. {\bf B400} (1993) 463.
\bibitem{modssym} A. Ceresole, R. D'Auria and S. Ferrara,
\Journal\PLB {339} {71}{1994}, hep-th/9408036.
\bibitem{FerStroCand} S.~Ferrara and A.~Strominger, in {\em Strings
'89}, eds. R.~Arnowitt, R.~Bryan, M.J.~Duff, D.V.~Nanopoulos and
C.N.~Pope (World Scientific, Singapore, 1989), p.~245;\\
P.~Candelas and X.~C.~de~la~Ossa, \Journal\NPB {355} {455}{1991},\\
P.~Candelas, X.~C.~de~la~Ossa, P.~Green and
L.~Parkes, \Journal\PLB{258}{118}{1991}; \Journal\NPB{359} {21}{1991}.
\bibitem{SeiWit} N.~Seiberg and E.~Witten, Nucl. Phys. {\bf
B426} (1994) 19; {\bf B431} (1994) 484.
\bibitem{grifharr} P.~Griffiths and J.~Harris, {\em Principles of algebraic geometry},
(J.~Wiley and Sons, 1951).
\bibitem{griffiths} P. Griffiths, Ann. Math. {\bf 90} (1969) 460,
496; \\
W. Lerche, D. Smit and N. Warner, Nucl. Phys. B372 (1992) 87.
\bibitem{klemmlerche} A.~Klemm, W.~Lerche, S.~Yankielowicz and S.~Theisen,
Phys. Lett. {\bf B344}  (1995) 169.
\bibitem{sundborg} U.~Danielsson and B.~Sundborg,
Phys. Lett. {\bf B358}  (1995) 273; hep-th/9504102;\\
A. Brandhuber and K. Landsteiner, Phys. Lett. {\bf B358} (1995) 73;
hep-th/9507008.
\bibitem{stefanlic} S. Rummens, licentiaatsthesis, K.U. Leuven, 1996.
\bibitem{rigidSUSYYM}
E. Martinec and N. Warner, Nucl. Phys. {\bf B459} (1996) 97;
hep-th/9509161;\\
P.C. Argyres and A.D. Shapere, Nucl. Phys. {\bf B461} (1996) 437;
hep-th/9509175;\\
A. Hanany, Nucl. Phys. {\bf B466} (1996) 85; hep-th/9509176;\\
U.~Danielsson and B.~Sundborg,
Phys. Lett. {\bf B370}  (1996) 83; hep-th/9511180;\\
M. Alishahiha, F. Ardalan and F. Mansouri, Phys. Lett. {\bf B381} (1996)
446; hep-th/9512005;\\
M.R. Abolhasani, M. Alishahiha and A.M. Ghezelbash, Nucl. Phys. {\bf
B480} (1996) 279; hep-th/9606043\\
W. Lerche and N.P. Warner, hep-th/9608183;\\
K. Landsteiner, J.M. Pierre and S.B. Giddings,
 Phys. Rev. {\bf D55} (1997) 2367; hep-th/9609059.
\bibitem{revN2} B. de Wit, in {\em Supergravity '81}, eds. S.
Ferrara and J.G. Taylor (Cambridge Univ. Press, 1982);\\
 A. Van Proeyen,
in {\em Supersymmetry and Supergravity 1983}, ed. B. Milewski, (World
Scientific, Singapore, 1983).
\bibitem{LondonN2sg} A. Van Proeyen, to be published
in the proceedings of the workshop "Gauge
  Theories, Applied Supersymmetry and Quantum Gravity", London Imperial
  College, july 1996; hep-th/9611112.
\bibitem{dWLVP} B. de Wit, P. Lauwers and A. Van Proeyen, Nucl.
Phys. {\bf B255} (1985) 569.
\bibitem{BaggerWittenN1} E. Witten and J. Bagger
\PLB{115}{202}{1982}; J. Bagger, in {\em Supersymmetry}, (NATO
   Advanced Study Institute, Series B: Physics, v. 125),
ed. K. Dietz et al., (Plenum Press, 1985).
\bibitem{Chern}
 S.-s.~Chern, Complex Manifolds Without Potential Theory (Springer,
1979).
\bibitem{Wells} R.~O.~Wells, Differential Analysis on Complex Manifolds
(Springer, 1980).
\bibitem{N=2N=1}  S. Ferrara, L. Girardello and M. Porrati,
\Journal\PLB{366}{155}{1996}; hep-th/9510074\\
P. Fr\`e,  L. Girardello, I. Pesando and M. Trigiante,
hep-th/9607032; and to be published
in the proceedings of the workshop "Gauge
  Theories, Applied Supersymmetry and Quantum Gravity", London Imperial
  College, july 1996; hep-th/9611188.
\bibitem{mincoupl} J.F. Luciani, Nucl. Phys. {\bf B132} (1978) 325;\\
M. de Roo, J.W. van Holten, B. de Wit and A. Van Proeyen, Nucl.
Phys. {\bf B173} (1980) 175\\
B. de Wit, J.W. van Holten and A. Van Proeyen, Nucl. Phys. {\bf B184}
(1981) 77 (E: {\bf B222}(1983) 516).
\bibitem{Seiberg} N. Seiberg, Nucl. Phys. {\bf B303} (1988) 286.
\end{thebibliography}
\end{document}